\documentclass[aps,prb,amsmath,amssymb,twocolumn,floatfix,showpacs,superscriptaddress]{revtex4-1}

\usepackage{graphicx}
\usepackage[usenames,dvipsnames]{color}
\usepackage{hyperref}
\usepackage{epstopdf}
\usepackage{times}
\usepackage{multirow}

\newcommand{\tr}{{\rm Tr}}

\newcommand{\Renyi}{R\'enyi\ }

\newcommand{\citationtitle}[1]{{\color{Gray}\small #1}}

\begin{document}

\title{A \Renyi entropy perspective on topological order in classical toric code models}
\author{Johannes Helmes}
\affiliation{Institute for Theoretical Physics, University of Cologne, 50937 Cologne, Germany}
\author{Jean-Marie St\'ephan}
\affiliation{Max-Planck-Institut f\"ur Physik komplexer Systeme, 01187 Dresden, Germany}
\author{Simon Trebst}
\affiliation{Institute for Theoretical Physics, University of Cologne, 50937 Cologne, Germany}
\date{\today}

\begin{abstract}
Concepts of information theory are increasingly used to characterize collective phenomena in condensed matter systems, such as the use of entanglement entropies to identify emergent topological order in interacting quantum many-body systems. Here we employ classical variants of these concepts, in particular \Renyi entropies and their associated mutual information, to identify topological order in classical 
systems. Like for their quantum counterparts, the presence of topological order can be identified in such classical systems  via a universal, subleading contribution to the prevalent volume and boundary laws of the classical \Renyi entropies.
We demonstrate that an additional subleading $O(1)$ contribution generically arises for all \Renyi entropies $S^{(n)}$ with $n \geq 2$ when driving the system towards a phase transition, e.g. into a conventionally ordered phase. 
This additional subleading term, which we dub connectivity contribution, tracks back to partial subsystem ordering and is proportional to the number of connected parts in a given bipartition. 
Notably, the Levin-Wen summation scheme -- typically used to extract the topological contribution to the \Renyi entropies -- does not fully eliminate this additional connectivity contribution in this classical context. This indicates that the distillation of topological order from \Renyi entropies requires an additional level of scrutiny to distinguish topological from non-topological $O(1)$ contributions. This is also the case for quantum systems, for which we discuss which entropies are sensitive to these connectivity contributions.
We showcase these findings by extensive numerical simulations of a classical variant of the toric code model, for which we study the stability of topological order in the presence of a magnetic field and at finite temperatures from a \Renyi entropy perspective.
\end{abstract}

\maketitle

%
%

\section{Introduction}

The ground states of interacting many-body systems can exhibit subtle forms of order such as the formation of long-range topological order or the emergence of spin liquid physics \cite{Balents} -- both in the quantum and classical world. Distilling the precise nature of these unconventional forms of order or even identifying their mere existence from, for instance, the ground state wavefunction of a quantum many-body system is a highly non-trivial endeavor. This is particularly true if one tries to rely on conventional characterization approaches (often inspired by their experimental feasibility) such as the use of correlation functions. As it is oftentimes the case, the breakthrough to overcome these limitations has come by completely changing one's point of view -- in this case by adopting an information theory perspective on the many-body system \cite{AreaLaw}.

One core concept of quantum information theory is to consider the entanglement for a bipartition of the many-body system into two parts and to precisely map out its dependencies on the size and shape of the boundary separating the two parts. This idea has been pioneered by Bekenstein and Hawking \cite{BekensteinHawking} who in their description of black holes have introduced the entanglement entropy as a quantitative measure of entanglement and demonstrated its characteristic boundary law scaling.
The best-known example of an entanglement entropy is the von-Neumann entropy
\begin{equation}
	S(A) = - \tr \left( \rho_A \ln \rho_A \right) \,,
	\label{eq:vonNeumann}
\end{equation}
where $\rho_A$ is the reduced density matrix of subsystem $A$ after tracing out subsystem $B$. By considering higher moments of the density matrix, one can embed the von-Neumann entropy into a family of \Renyi entropies
\begin{equation}
	S_n(A) = \frac{1}{1-n}  \ln \left( \tr \left[ \,\rho_A^n \, \right] \right) \,,
\end{equation}
where the index $n$ typically is an integer $n \geq 2$ with the limit $n\to 1$ recovering the von-Neumann entropy \eqref{eq:vonNeumann}.

\begin{figure}[t]
  \centering
  \parbox{0.21\linewidth}{\includegraphics[width=\linewidth]{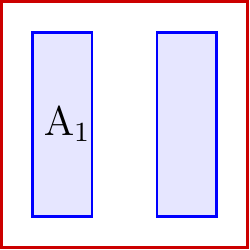}}
  \hspace{0.035\linewidth}
  \parbox{0.21\linewidth}{\includegraphics[width=\linewidth]{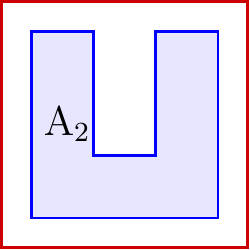}}
  \hspace{0.035\linewidth}
  \parbox{0.21\linewidth}{\includegraphics[width=\linewidth]{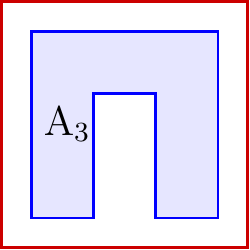}}
  \hspace{0.035\linewidth}
  \parbox{0.21\linewidth}{\includegraphics[width=\linewidth]{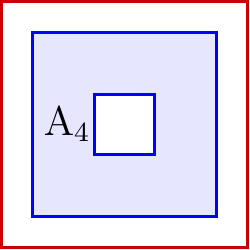}}
  \caption{(color online) Illustration of the four bipartitions used in the Levin-Wen summation scheme to extract the topological entropy.
  				     The latter is calculated
				     as $\gamma = -S(A_1) +S(A_2) +S(A_3) -S(A_4)$.}
\label{fig:LW}
\end{figure}

In contrast to thermal entropies these entanglement entropies are not extensive, but show a prevalent scaling with the length $\ell$ of the boundary of the bipartition \cite{AreaLaw}
\begin{equation}
        S_n(A) = a\cdot\ell + b \cdot \ln \ell - \gamma + O(1/\ell) \,,
        \label{eq:BoundaryLaw}
\end{equation}
where in addition to the boundary law scaling of the first term we explicitly list the first two subleading terms. 
The logarithmic corrections typically arise from corner contributions for gapless phases or quantum critical points \cite{MetlitskiGrover}.
We will focus in the following on the $O(1)$ contribution $\gamma$ that is typically associated with the topological entanglement entropy \cite{LevinWen,KitaevPreskill} for gapped two-dimensional phases, though such $O(1)$ contributions are also known to arise at quantum critical points reflecting geometrical aspects of the subsystem \cite{MetlitskiGrover}. For a gapped system a finite value of $\gamma$ not only indicates the presence of topological order, but also allows to narrow down \cite{FootnoteAmbiguities} the precise type of the underlying topological quantum field theory due to its universal character \cite{LevinWen,KitaevPreskill}. 
One way to calculate the entanglement entropy $\gamma$ from Equation \eqref{eq:BoundaryLaw} is to perform a careful scaling analysis for bipartitions of varying size, as it was done, for instance, in identifying the emergence of topological order in the ground state of the Heisenberg antiferromagnet on the kagome lattice \cite{JiangBalents,Depenbrock}. Alternatively, one can compute the topological entanglement entropy using one of the summation schemes put forward by Levin and Wen \cite{LevinWen} or Kitaev and Preskill \cite{KitaevPreskill} that consider a set of bipartitions of a system of fixed system size constructed in a way that all leading boundary and subleading corner contributions to the entanglement entropy precisely cancel. An illustration of the four bipartitions used in the Levin-Wen scheme is provided in Fig.~\ref{fig:LW}.

A similar information theory perspective can also be taken to identify subtle orders in {\em classical} many-body systems, which will be the subject of this manuscript. In making this quantum to classical transition, we follow the work of Castelnovo and Chamon \cite{castelnovo07_1} and consider the Shannon entropy
\begin{equation}
	S(A) = - \sum_{\{j_A\}} p_{j_A} \ln p_{j_A} \,,
	\label{eq:Shannon}
\end{equation}
where $\{j_A\}$ denotes the set of possible configurations of subsystem $A$ in a bipartition of the system with statistical weights $p_{j_A}$. 
More generally, we can consider an entire family of classical \Renyi entropies
\begin{equation}
	S_n(A) = \frac{1}{1-n} \ln \left( \sum_{\{j_A\}} p_{j_A}^n \right)  \,,
	\label{eq:clRenyi}
\end{equation}
where $n$ again is a positive integer $n\geq 2$ and the limes $n\to 1$ recovers the Shannon entropy \eqref{eq:Shannon}.

One key distinction of the classical versus quantum \Renyi entropies is that the classical \Renyi entropies follow a volume law
\begin{equation}
        S_n(A) = c_V \cdot V_A + c_\ell \cdot \ell - \gamma + O(1/\ell) \,,
        \label{eq:VolumeLaw}
\end{equation}
where $V_A$ now indicates the volume of partition $A$. The subleading $O(1)$ contribution $\gamma$ again indicates the existence of topological order. For the classical toric code, which we will discuss in more detail in the following, its value of $\gamma=\ln 2$ resembles the result of the quantum version up to a factor of 2 as first discussed 
in Ref.~\onlinecite{castelnovo07_1} and recently expanded to a family of more general classical stringnet models \cite{hermanns14}.

In this manuscript, we carefully analyze this $O(1)$ contribution to the classical \Renyi entropies when driving a system through a phase transition. As one particular example we consider the continuous phase transition from a topologically ordered to a conventionally ordered phase, which in the classical toric code can be driven by a magnetic field $h$. One of our main results is that for all higher-order \Renyi entropies $S^{(n)}$ with $n \geq 2$ there is an additional constant term beyond the topological contribution for the intermediate coupling regime $h_c/n \leq h \leq h_c$. This additional contribution arises from partial subsystem ordering in this coupling regime. Because it is sensitive to the number of disconnected parts of these subsystems we dub it a ``connectivity contribution" to sharply distinguish it from the known topological contribution. To illustrate the generic, non-topological nature of this $O(1)$ connectivity contribution we also demonstrate its occurrence for the two-dimensional Ising model driven through its thermal phase transition. Finally, we come back to the classical toric code and round off our discussion by analyzing the finite-temperature behavior of the $O(1)$ contributions to classical \Renyi entropies in a variety of settings.

The manuscript is organized as follows. We start with an introduction to the classical toric code in Sec.~\ref{sec:ClassicalToricCode}. We then turn to the information theory perspective of classical many-body systems and (re)introduce classical \Renyi entropies and related measures such as the mutual information and discuss their scaling behavior in Sec.~\ref{sec:ClassicalRenyiEntropies}. In this section, we also motivate and discuss our main results from an analytical perspective. This is followed by a discussion of extensive numerical simulations in Sec.~\ref{sec:numerics}. We conclude with an outlook on the applicability of our results to quantum many-body systems and a discussion of the general shortcomings of \Renyi entropies in Sec.~\ref{sec:outlook}. 
The main body of the manuscript is complemented by an extensive appendix, which gives a detailed exposition of the implementation of the replica technique to calculate the \Renyi entropies. In particular, we comprehensively discuss how to combine non-local loop update techniques with the replica scheme. 

%
%

\section{Classical toric code model}
\label{sec:ClassicalToricCode}

To set the stage for our analysis we give a short introduction to the classical toric code model, which will serve as a paradigmatic example of a classical system with topological order in our subsequent discussion. The classical toric code is derived from its well known quantum counterpart~\cite{kitaev03}, which is defined by the Hamiltonian
\begin{equation}
H = - J_p \sum_{p \in P} \prod_{i=1}^4 \sigma^x_{p_i}- J_v \sum_{v \in V} \prod_{i=1}^4 \sigma^z_{v_i} \,,
\end{equation}
where SU(2) spin-1/2 degrees of freedom located on the bonds of a square lattice are interacting via  four-spin interactions around the plaquettes $P$ and vertices $V$ of the lattice. The two (commuting) terms energetically favor closed loop configurations either in the $\sigma_x$- or $\sigma_z$-basis (which are dual to each other). The ground state of the Hamiltonian can then be written as a {\em loop gas}, i.e. an equal-weight superposition of all closed loop configurations (in one of the two bases). We will work in the $\sigma_z$-basis in the following.

The classical toric code is defined by interpreting the loop gas as a classical partition function, i.e. a partition sum to which all closed loop configurations contribute equally. We can define a Hamiltonian for this classical system in terms of Ising variables located on the bonds of a square lattice as
\begin{equation}
  H = - J_v \sum_{v \in V} \prod_{i=1}^4 \sigma_{v_i} \,,
  \label{eq:hamiltonian}
\end{equation}
where a four-spin term for each vertex again favors even parity spin alignment corresponding to closed loop configurations.
The low-temperature manifold of states for this classical Hamiltonian is thus again providing us with a loop gas. Like in the quantum model we can define four distinct winding number parity sectors, which for a toroidal geometry (periodic boundary conditions) give rise to a 4-fold degeneracy at zero temperature -- a hallmark of topological order also in the classical context. In the following, we will initially focus on this zero-temperature physics and impose a hard-constraint on the system enforcing closed loop configurations.

We can drive the classical toric code out of its topologically ordered phase by adding an external field term $- h \sum_{i=1}^N \sigma_i$ to the Hamiltonian that breaks the equal-weightness of the closed loop configurations constituting the loop gas. States with a higher total magnetization $m = \sum_{i=1}^N \sigma_i$ are preferred via statistical weights $e^{hm}$ (in lieu of uniform weights 1 for the unperturbed loop gas). Since a majority of spins pointing up in the toric code lattice effectively shortens the length of the individual loops, we will also call $h$ a \textit{loop tension}. For sufficiently large strength this loop tension will drive a transition from the loop gas to a conventionally polarized state where all spins are aligned along the external field. 

Note that in defining these statistical weights we have not introduced the notion of temperature. We will return to this point in Sec.~\ref{sec:finite-temperature} where we discuss the role of open loop defects arising at finite temperatures in Hamiltonian \eqref{eq:hamiltonian} and their interplay in the presence of a magnetic field.

\subsection{Kramers-Wannier duality and the 2D Ising model}
\label{sec:kramerswannier}

\begin{figure}[t]
  \centering
  \parbox{0.41\linewidth}{\bf Toric Code \\[0.7em] \includegraphics[width=\linewidth]{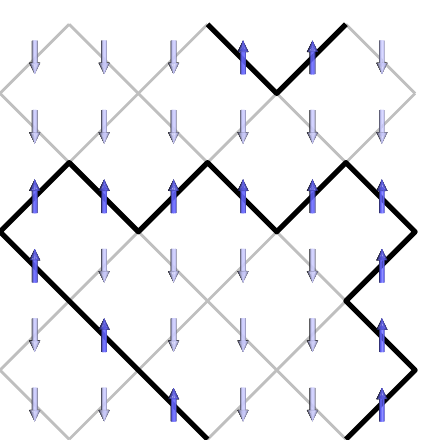}}
  \hspace{0.1cm}{\large $\longleftrightarrow$}\hspace{0.1cm}
  \parbox{0.41\linewidth}{\bf Ising model \\[0.7em] \includegraphics[width=\linewidth]{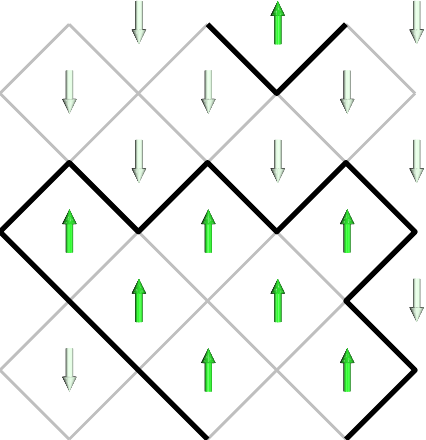}}
  \caption{(color online) The well-known Kramers-Wannier duality \cite{KramersWannier} maps the classical toric code in a magnetic field 
  		to the 2D Ising model at finite temperatures.}
\label{fig:mapping}
\end{figure}

The zero-temperature physics of the classical toric code in a magnetic field can be mapped via a Kramers-Wannier duality \cite{KramersWannier} to a 2D Ising model at finite temperatures -- much like its quantum counterpart in a magnetic field can be mapped to a 3D Ising model at finite temperatures \cite{Trebst07}.
To illustrate this mapping consider an alternative set of Ising degrees of freedom $\mu_i$ located at the center of the plaquettes $P$ of the original toric code model, see Fig.~\ref{fig:mapping}.  For a given loop configuration $\{\sigma\}$ the values of these alternate Ising spins are chosen such that $\mu_i \mu_j = -\sigma_{\langle i,j\rangle}$ where site $i$ and $j$ are nearest neighbors and $\sigma_{\langle i,j\rangle}$ is the spin on the toric code bond that separates plaquettes $i$ and $j$. The magnetic field term of the toric code model thus induces an Ising interaction between the dual degrees of freedom. Note that loops in the toric code correspond to domain walls in the Ising model as illustrated in Fig.~\ref{fig:mapping}.
 
This duality between the toric code  and the  Ising model readily implies that the critical value of the loop tension $h_c$ is directly related to the analytically known \cite{Onsager} critical value $\beta_c$ of the finite-temperature transition in the Ising model
\[
	h_{c}^{\text{TC}} = \beta_{c}^{\text{Ising}} = \ln(1+\sqrt2) / 2 \approx 0.440686 \ldots \,.
\]
It is also obvious that the mapping is unique up to a global flip of the Ising spins. In other words, the $\mathbb{Z}_2$ symmetry of the Ising model is lost in the toric code model.\cite{FootnoteBoundaryConditions}

The mapping also lets us understand the phase transition in the classical toric code as a continuous phase transition (in the 2D Ising universality class). In principle, one can of course track this transition by measuring, for instance, the total magnetization $m$ which serves as order parameter in the perspective of the Ising model. However, here we want to pursue a different direction and ask how other global quantities such as the \Renyi entropies can be used to quantitatively track this phase transition.

%
%

\section{Classical \Renyi entropies}
\label{sec:ClassicalRenyiEntropies}

With the classical toric code at hand, we now direct our analysis towards the signatures of topological order in classical \Renyi entropies. In this section our emphasis will be mostly on an analytical perspective. To this end, we first provide a short introduction to the replica description of \Renyi entropies. We then discuss the $O(1)$ contributions arising for the toric code model, in particular when driving the system through the field-induced phase transition to a topologically trivial phase. 


\subsection{Replica technique}
\label{sec:replica}

Let us start our more detailed discussion of \Renyi entropies by briefly laying out how one can calculate the \Renyi entropies using the so-called {\em replica technique}~\cite{Holzhey94}. The essential idea of this approach is to calculate the $n$-th \Renyi entropy from $n$ replicas of the representation of the underlying many-body system. While bipartition $B$ is allowed to independently fluctuate across the $n$ replicas, the fluctuations of bipartition $A$ are constrained to meet certain boundary conditions between the replicas.
This scheme was original developed in the context of analytical calculations \cite{Holzhey94} of \Renyi entropies for quantum field theories (on an $n$-sheeted Riemann surface) and later adapted to the analysis  quantum many-body systems in condensed matter physics \cite{Calabrese04}. It is also directly amenable to numerical simulations and has first been ported to Monte Carlo simulations in the context of  lattice gauge theories \cite{LatticeGauge}. For quantum lattice models the replica technique has first been reformulated in the context of stochastic series expansions in Ref.~\onlinecite{Hastings10}. Subsequently the replica technique has been adapted to various other flavors of quantum Monte Carlo including
   variational Monte Carlo \cite{ReplicaTrickVariational},
   continuum-space path integral Monte Carlo \cite{ReplicaTrickContinuum},
   determinantal Monte Carlo \cite{ReplicaTrickDQMC},
   fermionic continuous-time path integral Monte Carlo \cite{ReplicaTrickCT-QMC}, and
   hybrid Monte Carlo \cite{ReplicaTrickHybrid}.
More recently implementations of this replica technique have been envisioned for experimental settings \cite{experimental_renyi_2012},
which has led to the first experimental measurement of \Renyi entropies in an ensemble of ultracold atoms \cite{experimental_renyi_2015}.

\begin{figure}[b]
  \centering
  \includegraphics[width=\linewidth]{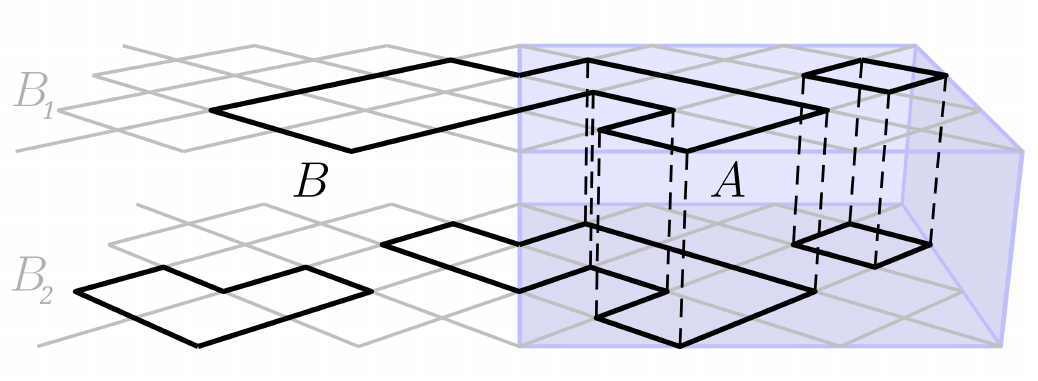}
  \caption{Illustration of the replica representation of the second \Renyi entropy $S_2(A)$.
  	       Pictured are two replicas of the toric code model with a possible loop gas configuration superimposed.
	       Similar to Fig.~2 the thick black lines represent up-pointing spins on the bonds of the lattice, while the down-pointing spins 
	       are not illustrated. Note that the loop configuration in the connected part $A$ is identical in both replicas, while two independent
	       loop configurations are allowed to occur in the two replicas of part $B$.}
  \label{fig:replicaloopgas}
\end{figure}

Here we follow the work \cite{Iaconis13} by Iaconis and collaborators and adapt the replica technique to {\em classical} systems -- an approach, which is again directly amenable to Monte Carlo simulations. We focus on the second \Renyi entropy $S_2(A)$ and therefore consider two replicas of the system where the degrees of freedom in part $A$ are precisely mimicking each other in the two replicas, while those in part $B$ are allowed to fluctuate independently for the two replicas -- see Fig.~\ref{fig:replicaloopgas} for an illustration.
This replica representation of the \Renyi entropy  
follows directly from writing
the second order \Renyi entropy via its definition (\ref{eq:clRenyi}) and employing statistical weights $\exp(hm)$ 
to capture the effect of a finite magnetic field of strength $h$
\begin{eqnarray}
  S_2(A) & = & - \ln \left[\sum_{\sigma_A, \sigma_B, \sigma_{B'}} \left(p_{\sigma_A\cup \sigma_B} \right) \left(p_{\sigma_A\cup \sigma_{B'}} \right)\right] \nonumber \\
  & = & - \ln \left[\sum_{\sigma_A, \sigma_B, \sigma_{B'}} \left( \frac{e^{h(m_A + m_B + m_A + m_{B'})}}{Z[h]^2} \right )\right] \nonumber \\
  & \equiv & - \ln \left[\frac{Z[A,2,h]}{Z[h]^2}\right] \,.
  \label{eq:reptrick}
\end{eqnarray}
Here we have introduced the partition function $Z[h]$ of the total system in the presence of a magnetic field of strength $h$. The second partition function $Z[A,2,h]$ captures a setting in which subsystem $A$ is subject to a magnetic field of strength $2h$, i.e. {\em twice} the strength of the applied magnetic field, while the two replicas for subsystem $B$ are subject to a magnetic field of strength $h$.

This replica representation of arbitrary \Renyi entropies with $n\geq2$ is then obtained in a straight-forward manner. In general, one  encounters partition functions of the form $Z[A,n,h]$ describing a system, in which subsystem $A$ experiences a magnetic field of strength $n  h$, while being coupled to $n$ independent replicas of subsystem $B$, each experiencing a magnetic field of strength $h$.


\subsection{Topological entropy and connectivity contribution}
\label{sec:theory}

\begin{figure}[b]
  \centering
  \begin{tabular}{|r||c|c|c|c|}
    \multicolumn{1}{c}{} & 
    \multicolumn{1}{c}{\includegraphics[width=0.18\linewidth]{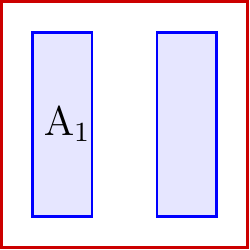}} & 
    \multicolumn{1}{c}{\includegraphics[width=0.18\linewidth]{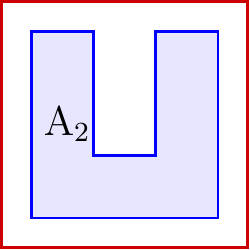}} & 
    \multicolumn{1}{c}{\includegraphics[width=0.18\linewidth]{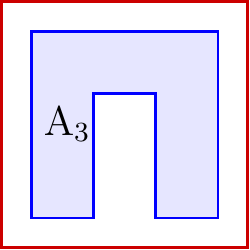}} & 
    \multicolumn{1}{c}{\includegraphics[width=0.18\linewidth]{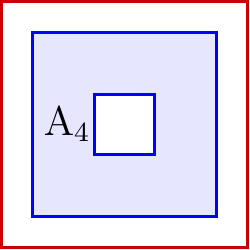}} \\ \hline
    \multicolumn{5}{|c|}{topological $(h<h_c/2)$} \\ \hline
     \parbox[c][2 em][c]{0.08\linewidth}{$\gamma^A$} & 0 & 0 & 0 & $\ln 2$ \\ \hline
     \parbox[c][2 em][c]{0.08\linewidth}{$\gamma^B$} & $\ln 2$ & 0 & 0 & 0  \\ \hline\hline
    \multicolumn{5}{|c|}{connectivity $(h_c/2 < h < h_c)$} \\ \hline
     \parbox[c][2 em][c]{0.08\linewidth}{$\gamma^A$} & $ 2\ln 2$ & $\ln 2$ &  $\ln 2$& $ 2\ln 2$ \\ \hline\hline
     \parbox[c][2 em][c]{0.08\linewidth}{$\gamma^B$} & $ 2\ln 2$ & $\ln 2$ &  $\ln 2$& $2\ln 2$ \\ \hline\hline
  \end{tabular}
  \caption{(color online) Constant contributions to the second \Renyi entropy in the topologically ordered phase of the classical toric code model in a magnetic field of strength $h$ for the different bipartitions in the Levin-Wen scheme. For simplicity $n=2$ is shown.}
  \label{fig:LevinWen}
\end{figure}

As mentioned already in the introduction, the classical \Renyi entropies generically follow a volume law as given in Eq.~\eqref{eq:VolumeLaw} augmented by a subleading boundary law and $O(1)$ contributions indicative of topological order. We will concentrate on precisely these $O(1)$ contributions in the following. To extract them from the \Renyi entropy we employ the Levin-Wen summation scheme \cite{LevinWen} considering the four bipartitions $A_1$, $A_2$, $A_3$ and $A_4$ illustrated in Fig.~\ref{fig:LW}. Adding up the \Renyi entropies of these four bipartitions as
\begin{equation}
\Delta S= -S_2(A_1) + S_2(A_2) + S_2(A_3) - S_2(A_4),
 \label{eq:LevinWendiff}
\end{equation}
we note that all volume, boundary, and possible corner contributions to the \Renyi entropies precisely cancel. Hence, $\Delta S$ provides us with a direct measurement of the remaining $O(1)$ contributions. Following the work of Ref.~\onlinecite{hermanns14}, we first note that in the classical context only the fourth bipartition in the Levin-Wen scheme contributes a non-zero {\em topological} contribution as listed in the Table of Fig.~\ref{fig:LevinWen}.
It is this non-vanishing contribution, which provides us with a topological entropy of precisely $\gamma = \ln 2$ for the unperturbed classical toric code as originally shown by Castelnovo and Chamon \cite{castelnovo07_1}.

We now consider the effect of a finite loop tension (or magnetic field). From the Kramers-Wannier duality of the Ising model we already know that only when the loop tension reaches the critical value $h_c=\ln(1+\sqrt{2})/2$ topological order in the system is destroyed. When considering the regime $h<h_c$ from a renormalization group perspective, the effect of a loop tension renormalizes to zero in the thermodynamic limit and this physics should also be reflected in the entropy. One might thus be (mis)led to expect a constant $\gamma = \ln 2$ contribution to the classical entropy all the way up to the critical value $h_c$.
 
In fact, however,  the situation turns out to be a bit more subtle for \Renyi entropies. To get the topological entropy, it is mandatory that {\em all} replicas remain topologically ordered. But as we have seen in the replica representation of the \Renyi entropies in Sec.~\ref{sec:replica}, the two subsystems are exposed to different strengths of the loop tension, e.g. for the second \Renyi entropy one has field strengths $2h$ and $h$ for subsystems $A$ and $B$, respectively. This requires to refine our (renormalization group) perspective to consider \emph{three} separate parameter regions. 
 
 When $h<h_c/2$, all replicas are in the topological phase. Here the loop tension is an irrelevant perturbation (in an renormalization group sense), and the arguments of Ref.~\onlinecite{castelnovo07_1} apply. Hence we expect a purely topological $O(1)$ contribution of $\gamma = \ln 2$ to be calculated by the Levin-Wen scheme.
 
 When $h>h_c$ all replicas are in the ordered phase, with most spins pointing up. As can be understood by considering the limiting case $h\to \infty$ with zero entropy, the four contribution of Eq.~(\ref{eq:LevinWendiff}) become identical and the $O(1)$ contribution $\Delta S$ vanishes in the thermodynamic limit.

 The intermediate region $h_c/2<h<h_c$ is somewhat more delicate. Here we have one replica $A$ already in the ordered phase, while coupled to $n$ replicas of $B$ still in the topological phase. In case the bipartition leads to a single boundary (i.e. bipartitions $A_2$ and $A_3$) this generates an extra contribution $\ln 2/(n-1)$ to the entropy. In case the bipartition gives two separate boundaries (i.e. bipartitions $A_1$ and $A_4$, in which either subsystem $A$ or $B$ comes in two disconnected parts) the contribution becomes $2\ln 2/(n-1)$. We name this contribution \emph{connectivity contribution}, as it has no topological origin and is instead sensitive to the disconnected parts in subsystem $A$ or $B$. 
 The various contributions to the entropy for each geometry are detailed in Table.~\ref{fig:LevinWen}, specifying the situation of $n=2$ for simplicity.

Adding up the individual contributions, the Levin-Wen summation scheme then yields
\begin{equation}
 \Delta S=\left\{
 \begin{array}{ccc}
  \ln 2&,&h<h_c/2\\
  \frac{n}{n-1}\ln 2&,&h_c/2<h<h_c\\
  0&,&h>h_c
 \end{array}
 \right.
\end{equation}
in the three different parameter regions.

This is one of the main results of this paper. It will be checked numerically for $n=2$ and large system sizes in Sec.~\ref{sec:numerics}. Let us finally comment on the two special values $h=h_c/2$ and $h=h_c$. In that case, one or several of the replicas are exactly at the critical temperature in the language of the dual Ising model. These replicas are now governed by the Ising conformal field theory. It is known \cite{CardyPeschel} that at the critical point the corner contributions are enhanced to logarithmic terms $\propto \ln (L/a)$, where $a$ is some microscopic cutoff of the order of a lattice spacing. As is explained in Ref.~\onlinecite{IsakovHastingsMelko}, these logarithmic terms are canceled by the linear combination (\ref{eq:LevinWendiff}). However, there is no reason for all microscopic cutoffs to be the same. This means there is going to be an extra non-universal $O(1)$ contribution to $\Delta S$ arising in the vicinity of the critical points. Hence we expect a non-vanishing, but non-universal value for $\Delta S$ exactly at the critical points $h=h_c/2$ and $h=h_c$.


%
%

\section{Numerical Simulations}
\label{sec:numerics}
 
To complement the analytical perspective on \Renyi entropies for the loop tension driven phase transition in the toric code we will now
turn to a discussion of extensive numerical simulations of this model. 


\subsection{Monte Carlo setup}

We start our discussion of the numerical simulations by first outlining some aspects of our classical Monte Carlo setup, in particular with regard to the choice of bipartition, the numerical adaptation of the replica technique as well as the implementation of non-local loop update techniques.

\subsubsection*{Bipartition}
 
\begin{figure}[t]
  \centering
  \includegraphics[width=0.75\columnwidth]{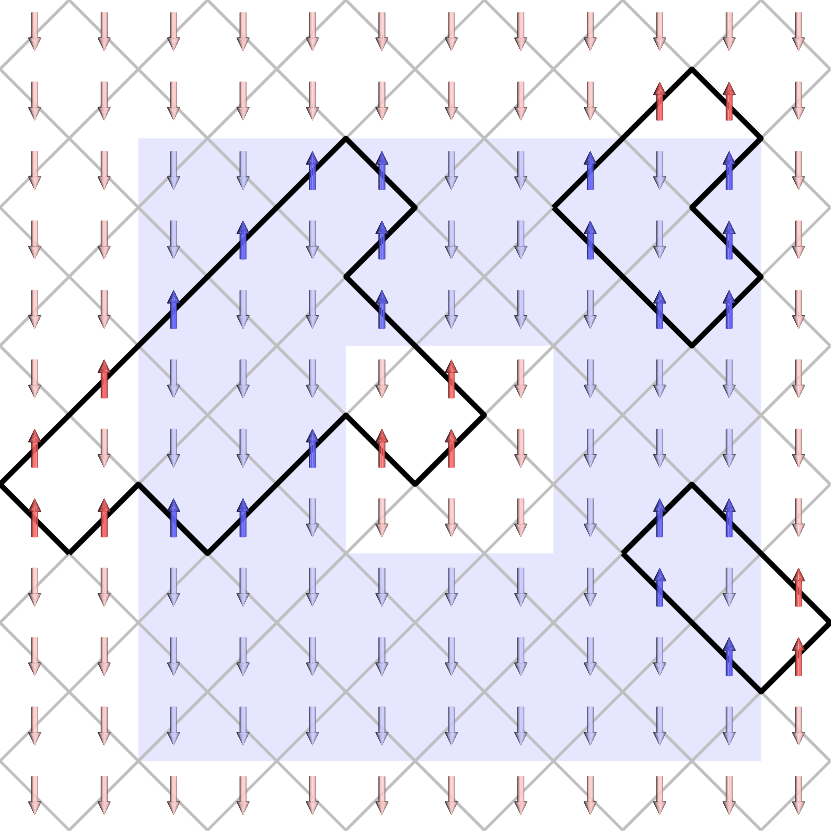}
  \caption{(color online) Fourth bipartition of the Levin Wen scheme as an example of how the subdivision is performed. Spins in subsystem $A$ are shown in blue, spins in $B$ are in red. The loops are represented by thick black lines.}
  \label{fig:bipartition}
\end{figure}

Let us first discuss our general choice for  bipartitioning the system. A naive approach would be to cut along the lines of the square lattice and therefore also cut through spins. This is certainly not desirable, as it would also imply that some loop segments could run along the boundary of the bipartition (see also Fig.~\ref{fig:mapping}) and in general it would leave us wondering how to assign the degrees of freedom on the boundary to the respective subsystems. 
Instead we choose to rotate the toric code lattice by 45 degrees and perform diagonal cuts in a way such that each plaquette is cut symmetrically along each line of the bipartition. This is illustrated for subsystem $A_4$ in Fig.~\ref{fig:bipartition}.

\subsubsection*{Replica technique}
We employ two alternative approaches to implement the replica representation \eqref{eq:reptrick} and calculate the \Renyi entropy from the ratio of the partition functions
$Z[h]$ and $Z[A,2,h]$ as 
\[
     S_2(A) = - \ln \left[\frac{Z[A,2,h]}{Z[h]^2}\right] 
\]
in a classical Monte Carlo setup. 

The first approach uses the fact that for a given partition function, e.g. $Z[h]$, we have $\frac{\partial \ln Z[h]}{\partial h} = \langle m \rangle$, where $\langle m \rangle$ is the average total magnetization. We can thus obtain the entropy by performing an integration over the estimated values of $\langle m \rangle$ for a dense range of parameters $h$. This method is an adaptation of the thermodynamic integration employed e.g. in the context of thermal phase transitions\cite{Melko10,Iaconis13}. 

 The second approach is often referred to as \textit{ensemble switching} \cite{Roscilde12} as it relates the ratio of the two partition functions $Z[h]^2$ and $Z[A,2,h]$ to the number of times one can switch between the two ensembles for a given sequence of configurations in a Markov chain. To this end, one samples configurations that belong to the system with partition function $Z[h]^2$, i.e. two copies of the full system. 
Since the configuration space of $Z[h]^2$ fully includes the one of $Z[A,2,h]$ we can readily estimate the relative weight of the two partition functions (needed for the above ratio) by simply counting how often a configuration in the Markov chain of the partition function $Z[h]^2$ also belongs to the one for partition function $Z[A,2,h]$. The ratio between the number of such coincidences and the total number of simulation steps is precisely $Z[A,2,h] / Z[h]^2$.

It turns out that neither of the two strategies is preferable for a wide range of system sizes. The thermodynamic integration needs a constant amount of individual simulations in order to densely cover the integration space, independently of the system size, while for the ensemble switching only a single simulation is needed. However, a common configuration of both $Z[h]^2$ and $Z[A,2,h]$ requires the full agreement of all spins in part $A$ and thus becomes a rare event when part $A$ is large leading to a dramatic increase in simulation times to obtain small statistical errors. This drawback can be partially cured by a so-called increment trick \cite{Hastings10}, which subdivides the problem into individual simulations for subsets of a partition of $A$. 

We find that thermodynamic integration typically outperforms the ensemble switching method for large systems (i.e. L $\gtrsim 48$), while the reverse is true for smaller sizes. As such we employ both approaches in the following.

\subsubsection*{Loop updates}

To efficiently sample the loop gas configurations constituting the partition function of the classical toric code non-local update techniques are an essential requirement. While such non-local update techniques, e.g. in the form of loop updates, are well known and widely employed, for instance, in the context of stochastic series expansions \cite{DirectedLoops}, the combination of these techniques with the replica technique has not been developed so far (neither in the context of quantum systems nor for classical loop models). The complication one faces when combining these two approaches is manifesting itself at the boundary of the bipartition where valid loop configurations must be sampled at all times. In particular, this implies that when iteratively building up the loop update the algorithm needs to branch out into multiple replicas, e.g. when building a loop starting from subsystem $A$ and branching out into the independent subsystems $B_1, B_2, \ldots B_n$, see Fig.~\ref{fig:replicaloopgas} for an illustration. One major technical achievement of this paper is the development of precisely such a (directed) loop update algorithm that {\em efficiently} samples replica-consistent loop gas configurations for closed loops as well as excited states with open loops. We refer to the appendix for a detailed description of this update scheme and only want to briefly sketch the main ideas in this section.

In an initial strategy we separate the tasks of identifying and accepting a new configuration. The first task is accomplished by performing an unbiased random walk that starts at an arbitrary vertex in the replicated lattice and continues through the lattice until it goes back to its initial vertex. If it enters the connected subsystem $A$ the random walk simultaneously traverses in both replicas and subsequently bifurcates at the boundary when transitioning back into the multiple independent replicas of subsystem $B$. If such bifurcations occur, we need to not only rejoin the first vertex, but also ensure that all open ends at these intermediate bifurcations are eventually turned into closed loop configurations. Once a valid closed loop structure has been identified the statistical weight of the resulting new spin/loop configuration (in which all spins are flipped for all segments along the identified loop structure) is determined and the resulting configuration is accepted with Metropolis probability. This update procedure can easily be adapted to a case where also open loop ends are allowed (a scenario relevant for  finite temperature simulations discussed in Section \ref{sec:finite-temperature}). The probability for the emergence of a new pair of defects is simply included in the Metropolis selection.

A more advanced strategy to maximize the acceptance probability of the new configuration is to introduce statistical weights already when building the update loop. To do so, the random walk which samples a new loop in the lattice is no longer unbiased but {\it directed}, i.e. the direction of turns in the loop update depends on the local loop configuration at every vertex and is biased with certain statistical weights. Note, that in our context the term {\it directed loop} refers to loops in the lattice and not in the imaginary-time expansion of world-line or SSE-type quantum Monte Carlo approaches. Also for this advanced strategy the update loop undergoes intermediate bifurcations if it exits subsystem $A$ and we have to perform the walk until all open ends (including those occurring at the boundary between the two subsystems) are rejoined. Allowing for open loop ends (e.g. in the  finite-temperature simulations of Section \ref{sec:finite-temperature}) this approach is readily modified by adding the possibility to stop the loop at a specific vertex and thereby creating an excitation.

For further details on these two approaches, in particular the detailed balance equations required to implement the directed loop updates, we refer to the extensive appendix of this manuscript.

To obtain our simulation results we used a combination of both algorithms since it turns out that the unbiased random walk strategy is more efficient for small loop tensions $h$, while the directed walk is performing better for intermediate to large loop tensions $h \gtrsim 0.3$.


\subsection{Mutual information}
\label{sec:minumerics}

We start our presentation of the numerical results by first looking at the (R\'enyi) mutual information -- a quantity that tracks long range correlations between two parts of a many-body system. It is obtained from the \Renyi entropies as
\begin{equation}
  I_2(A:B)=S_2(A) + S_2(B) - S_2(AB) \,.
  \label{eq:mi}
\end{equation}
Heuristically, the mutual information indicates how much information may be gathered about subsystem $A$ by doing all possible measurements in subsystem $B$ (and vice versa). Such an information retrieval about part $A$ occurs in particular in the presence of long-range correlations, which mediate some of the information about subsystem $A$ into subsystem $B$ where it can be gathered. As such, we expect the classical mutual information to be large in an ordered phase and easily distinguished from the one of a disordered phase where it should be considerably reduced. The mutual information has indeed proven to be a versatile indicator of phase transitions \cite{Iaconis13,GMI}; it is therefore natural to also consider it in the context of topological order as we will do in the following. 

Before doing do, we note that contrary to the classical entropy, the mutual information obeys a {\em boundary} law 
\begin{equation}
    I_2(A:B) = c_\ell \cdot (\ell-1) + a
    \label{eq:MI-scaling}
\end{equation}
as all bulk contributions are canceled by definition. $I_2$ is instead sensitive to subleading contributions to the \Renyi entropies, which we have specified here as a boundary and a constant contribution.
Focusing on the boundary contribution first, note that we wrote down a term proportional to $(\ell -1)$ instead of the total length $\ell$ of the boundary. The reason for this reduced scaling form originates directly from the loop gas constraint that all loops need to be closed. As such only an even number of loop segments is allowed to cross the boundary between the two subsystems $A$ and $B$, i.e. specifying the loop occupation of $\ell -1$ bond segments along the boundary we can readily deduce the occupation of the last bond segment. This argument not only provides us with the $\ell -1$ factor, but also readily determines the coefficient of the boundary contribution to be $c_\ell = \ln 2$. For further details of this argument we refer to Ref.~\onlinecite{hermanns14}, in which classical loop gas (and string net) states have been extensively discussed from a combinatorial perspective.
Let us now turn to the subleading constant contribution $a$ to the mutual information. This contribution can be readily calculated from the $O(1)$ contributions to the \Renyi entropies as specified in the table of Fig.~\ref{fig:LevinWen}. Noting that for a half-torus bipartition as employed in our numerical simulations (see below) no topological $O(1)$ contributions are expected to occur, we directly probe the connectivity contributions to the entropies. We thus expect the constant contribution to the mutual information to be
\begin{equation}\label{eq:miprediction}
 a=\left\{
 \begin{array}{ccc}
  0&,&h<h_c/n \nonumber \\
  -\frac{1}{n-1}\ln 2&,&h_c/n<h<h_c \nonumber \\
  0&,&h>h_c \,,
 \end{array}
 \right.
\end{equation}
for arbitary \Renyi entropies with $n \geq 2$.

\begin{figure}[t]
\includegraphics[width=\linewidth]{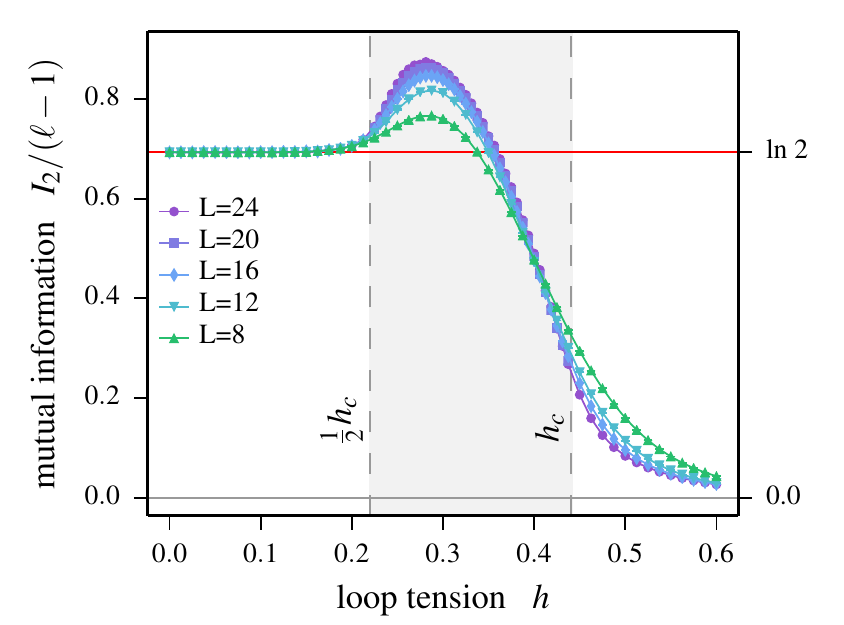}
\includegraphics[width=\linewidth]{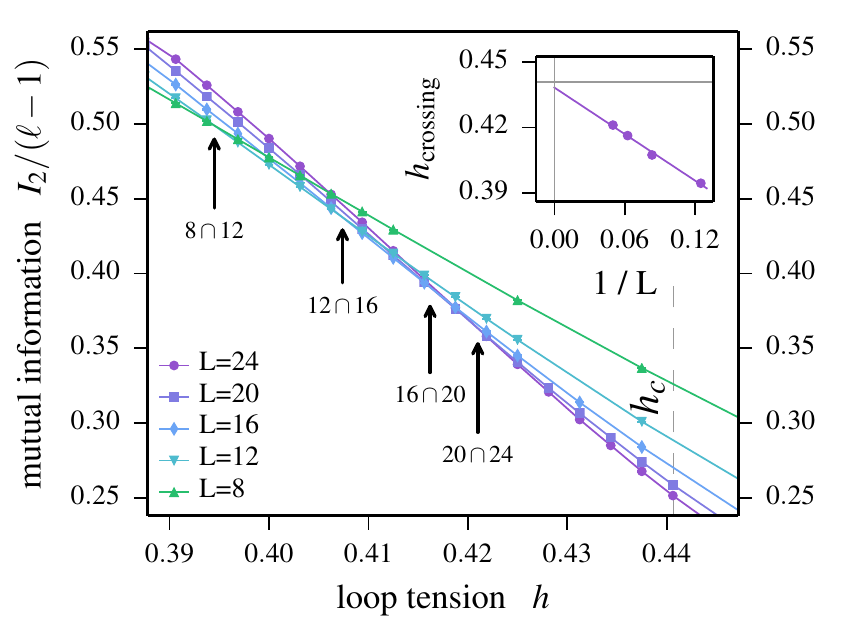}
\caption{(color online) 
 	       {\it Top panel:}  
	       	mutual information $I_2$ of the classical toric code model with a magnetic field/loop tension $h$ 
	       	for varying system sizes. The grey shaded area indicates the intermediate regime $h_c/2 \leq h \leq h_c$.
	       {\it Bottom panel:} 
	       Zoom of the data near $h_c$ illustrating the crossing points for different system sizes. 
	       The inset shows a scaling of the crossing points with inverse system size $1 / L$ extrapolating well
	       to the expected value of $h_c = \ln(1+\sqrt2) / 2 \approx 0.440686 \ldots $ in the thermodynamic limit 
		(indicated by the grey lines).
         }
\label{fig:mutual_inf}
\end{figure}

In our numerical setup for the classical toric code, we choose a symmetric bipartition of a quadratic lattice of size $L \times L$ with periodic boundary conditions (torus) where part $A$ is precisely one half of the torus. The boundaries are two straight lines of length $\ell = L$. The symmetry of this bipartition readily implies $S_2(A)=S_2(B)$. 
Our results for the mutual information $I_2(A:B)$ are plotted in 
Fig.~\ref{fig:mutual_inf} for varying values of the  loop tension $h$. Indeed the illustrated behavior of the mutual information -- rescaled by the length of the boundary -- nicely confirms our expectation: in the loop gas limit of small loop tension the data for different system sizes $L$ collapse onto a single curve, which saturates at the expected value of $c_\ell = \ln 2$. For strong loop tension $h>h_c$ where the system is in a polarized state the mutual information drops to zero as expected.
Of interest is the intermediate regime $h_c/2 \leq h \leq h_c$, in which the mutual infomation first splits at $h_c/2$ for different system sizes, overshoots the value of $\ln 2$ and subsequently exhibits a crossing points at $h_c$. The behavior in the intermediate region can be traced back to the fact that subsystem $A$ is effectively in the polarized state, which leads to an overestimation of the information to be gained about this state from the perspective of subsystem $B$. 

\begin{figure}[t]
\includegraphics[width=\linewidth]{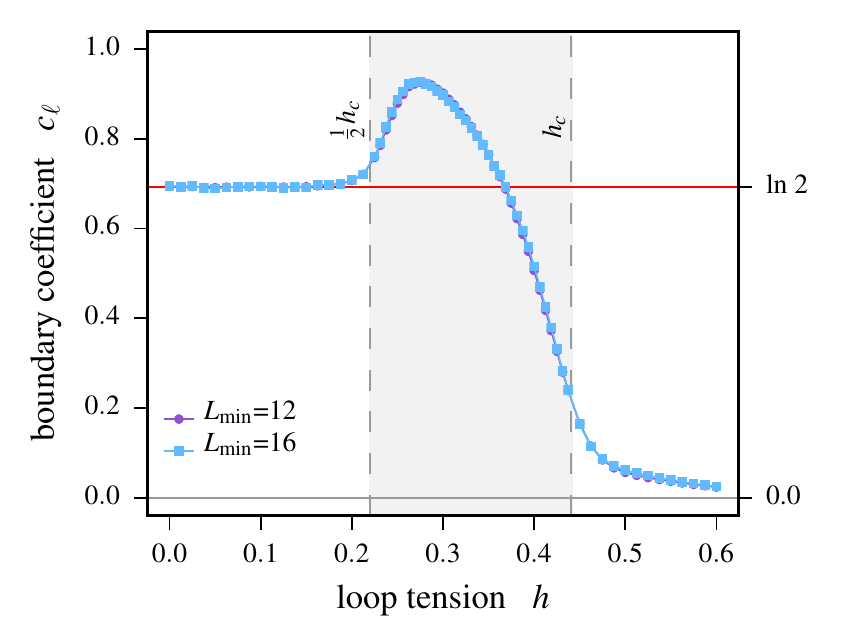}
\includegraphics[width=\linewidth]{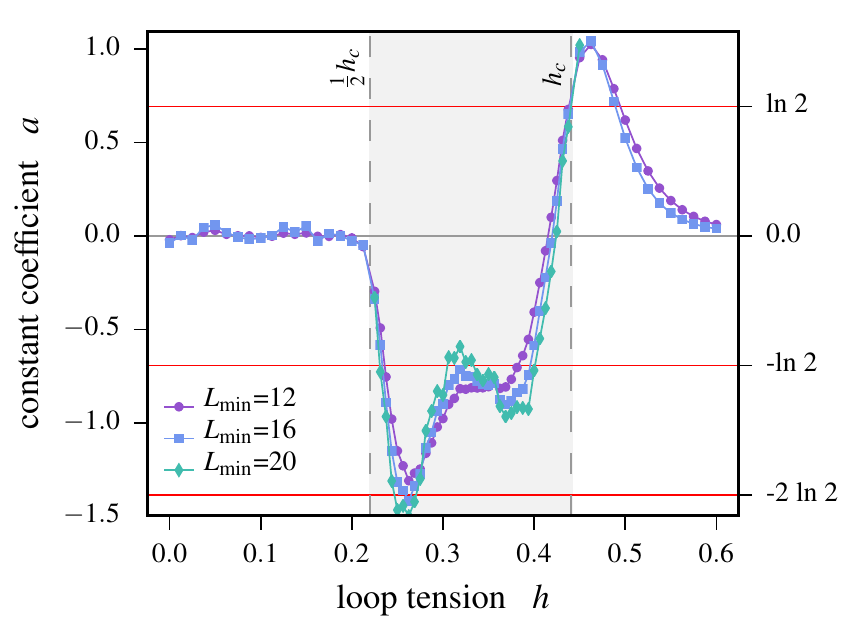}
\caption{(color online) 
	     Contributions to the finite-size scaling of the mutual information 
	     for the classical toric code model with magnetic field/loop tension $h$.
	     We fit the mutual information to the form $I_2(A:B) = c_\ell \cdot (\ell-1) + a$  
	     where $\ell$ is the length of the boundary of the bipartition.
	     The top panel shows the boundary coefficient $c_\ell$ 
	     and the bottom panel the constant contribution $a$ of this fit.
	     Data is obtained from fitting system sizes $L_{\rm min}=12$ to $L_{\rm max}=24$ 
	     and  $L_{\rm min}=16$ to $L_{\rm max}=24$, respectively.
	     The grey shaded area indicates the intermediate regime $h_c/2 \leq h \leq h_c$.
}
\label{fig:mutual_inf_fit}
\end{figure}

To understand why we see a splitting versus a crossing point at $h_c/2$ and $h_c$, respectively, it is worthwhile taking a look at the subleading contributions to the mutual information. To this end, we have fitted our numerical results for the mutual information for different system sizes to the expected scaling behavior of Eq.~ \eqref{eq:MI-scaling}.  Results for the obtained boundary coefficient $c_\ell$ and constant coefficient $a$ are shown in Fig. \ref{fig:mutual_inf_fit}. It is confirmed that the boundary coefficient $c_\ell$ attains a limiting curve which coincides with Fig.~\ref{fig:mutual_inf} in the thermodynamic limit. More interesting is the constant contribution $a$ whose sign changes explain the splitting vs. crossing points for different system sizes. In accordance with our expectation we see that this constant term does follow the behavior described in Eq.~\eqref{eq:miprediction}. In addition, we see that the coefficient $a$ undergoes a sign change at $h_c$, which results in the crossing point of the mutual information \cite{Iaconis13}. In contrast, at $h_c/2$ it does not change signs but goes from zero to negative, which in turn yields the splitting of the mutual information (but not a crossing point).


\begin{figure}[t]
\includegraphics[width=\columnwidth]{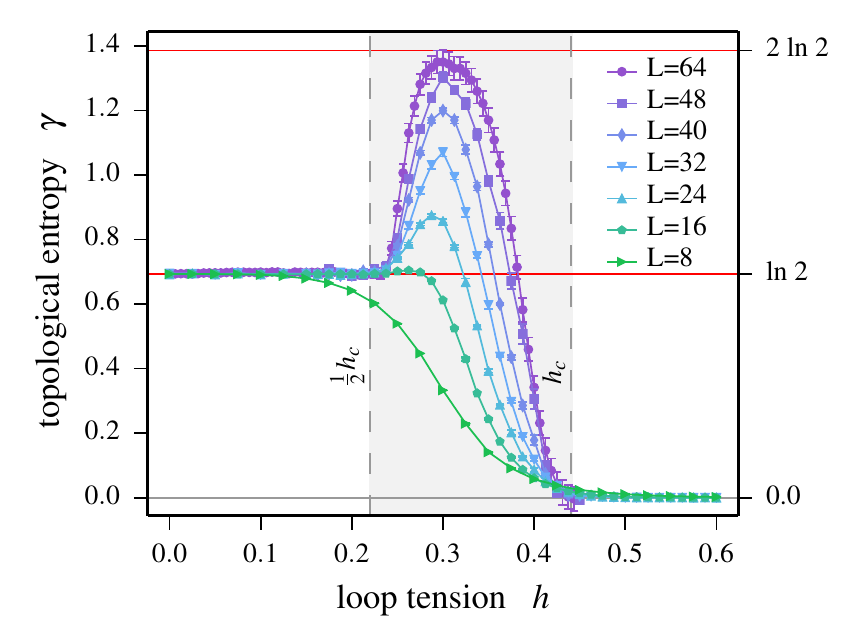}
\caption{(color online) 
	        The $O(1)$ contribution to the \Renyi entropy $S^{(2)}$ of the classical toric code model versus magnetic field/loop tension $h$
	        determined via the Levin-Wen summation scheme. The $O(1)$ contribution contains the expected universal topological 
		contribution of $\gamma = \ln(2)$ in the regime $h \leq h_c$ as well as an additional connectivity contribution $\gamma = \ln(2)$
		in the intermediate regime $h_c/2 \leq h \leq h_c$ (indicated by the grey shaded area), in which for sufficiently large system sizes 
		subsystem $A$ already transitions into the paramagnetic phase.
}
\label{fig:gamma_total}
\end{figure}

\begin{figure}[!h]
\includegraphics[width=\columnwidth]{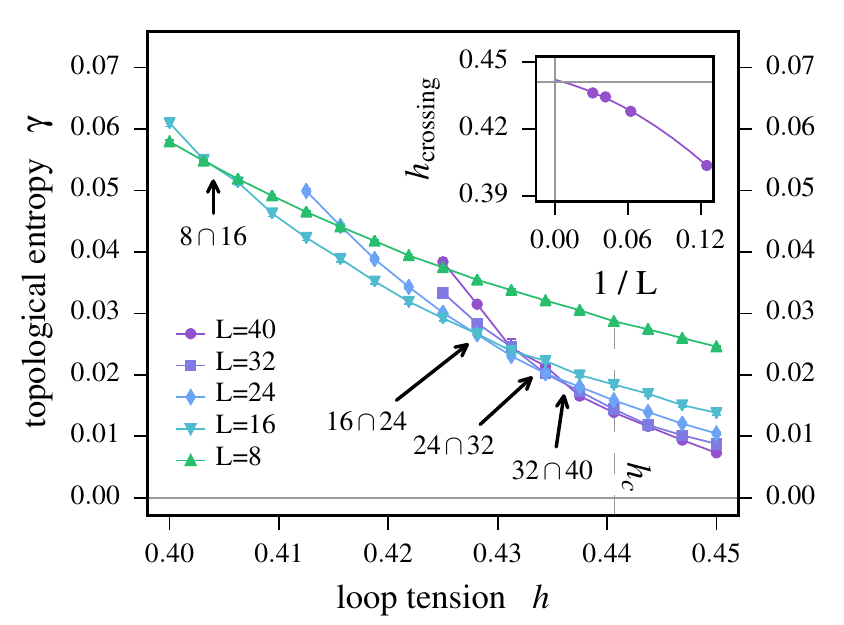}
\caption{(color online) Near the critical loop tension, crossing points between data curves of different linear system sizes $L$ can be identified. While keeping the difference between crossing $L$'s constant ($\Delta L=8$) a clear tendency towards the critical point can be observed upon increasing $L$. However, an extrapolation of the location of the crossing points is non-trivial because a subleading logarithmic contribution coming from corners of the subsystem is expected at criticality unlike in off-critical regions. 
}
\label{fig:gamma_critical}
\end{figure}

\subsection{Topological entropy and connectivity contribution} 

Our main quantity of interest is the classical topological entropy which we extract via the Levin-Wen scheme from the \Renyi entropy. In order to reduce effects from asymmetries in the bipartition we choose the thickness of any building blocks in the scheme to be $L/4$. We determine this entropy for a range of loop tensions $0 \leq h \leq 0.6$ since we expect it to reflect the phase transition at $h_c \approx 0.44$. For small system sizes we find a validation of the analytical expectation for the limits of $h \ll h_c$ and $h > h_c$ provided in Sec.~\ref{sec:theory} with a continuous decrease of the topological entropy between the limiting values of $\ln 2$ and $0$ around $h_c$, see Fig.~\ref{fig:gamma_total}. However, upon increasing $L$ an additional feature emerges in the range of $h_c / 2 < h < h_c$ -- an interim overshooting of $\gamma$ before it vanishes as expected near $h_c$. This overshooting converges to a plateau at a value of $2 \ln 2$ for large $L$.
As discussed in Sec.~\ref{sec:theory} this overshooting is an artifact of the \Renyi entropies with $n\geq2$ and arises from the connectivity contribution. Following the table of Fig. \ref{fig:LevinWen} we see that it yields a value of $2 \ln 2$ for $n=2$, to which the numerical data in Fig.~\ref{fig:gamma_total} nicely converges. 

\subsubsection*{$O(1)$ contributions for the 2D Ising model}

\begin{figure}[t]
\includegraphics[width=\linewidth]{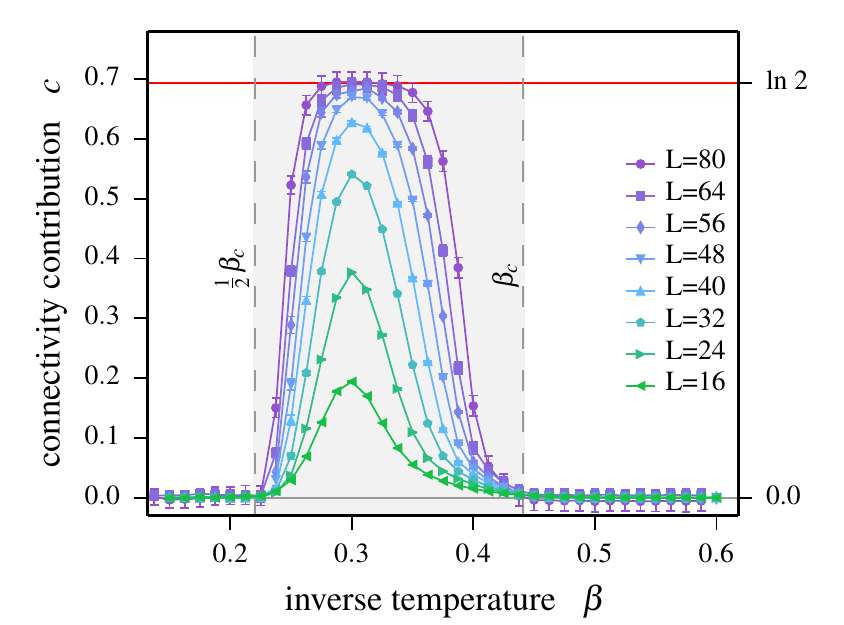}
\caption{(color online) 
	        The $O(1)$ contribution to the \Renyi entropy $S^{(2)}$ of the 2D Ising model versus inverse temperature $\beta$
	        determined via the Levin-Wen summation scheme. While no topological contribution is expected, a finite $O(1)$ 
	        connectivity contribution of $\gamma = \ln(2)$ is observed in the intermediate regime $\beta_c/2 \leq \beta \leq \beta_c$
	        (indicated by the grey shaded area),
	        in which for sufficiently large system sizes subsystem $A$ already transitions into the low-temperature ordered phase.
}
\label{fig:ising}
\end{figure}

The occurence of this contribution evokes the question if it also arises in a system {\em without} topological order. For this reason we consider the 2D Ising model on the square lattice and extract numerically the $O(1)$ contribution to the \Renyi entropy when driving the Ising model through its thermal phase transition.
Results are shown in Fig. \ref{fig:ising} where we see the formation of a plateau at $\ln 2$ in the intermediate temperature regime between $\beta_c / 2$ and $\beta_c$ upon increasing system size. In this intermediate region, subsystem $A$ already exhibits magnetic order. Since here we now allow for spontaneous symmetry breaking the two possible magnetically ordered states contribute a term $\ln 2$ per disconnected part of subsystem $A$, i.e. $2 \ln 2$ for bipartition $A_1$ and $\ln 2$ for bipartitions $A_2$, $A_3$, and $A_4$. Note that this is only the case if the disconnected parts of subsystem $A$ are spatially separated beyond the correlation length -- which explains the saturation on the plateau only for sufficiently large system sizes.

For higher \Renyi entropies with $n \geq 2$ we thus expect a plateau at a level of $\ln 2/(n-1)$ between $\beta_c /n$ and $\beta_c$ in the thermodynamic limit. 

\subsection{Finite-temperature simulations}
\label{sec:finite-temperature}

For finite-temperatures, excitations that break the closed loop constraint of the zero-temperature loop gas are thermally activated.
These thermally induced defects, i.e. open loops in the configurations, are expected to drive the system into a disordered phase
and hence destroy its topological order. In fact, in the thermodynamic limit it is known \cite{castelnovo07_2} that topological order does not survive for any finite temperature $T>0$. For finite system sizes, however, a thermal crossover between the topologically ordered and the disordered phase takes place at some finite crossover temperature $T_\text{co}(L)$. As pointed out in Ref.~\onlinecite{castelnovo07_2} the crossover temperature vanishes as $T_\text{co}(L) \sim (\ln L)^{-1}$, both for the quantum and classical toric code.

\begin{figure}
\includegraphics[width=\linewidth]{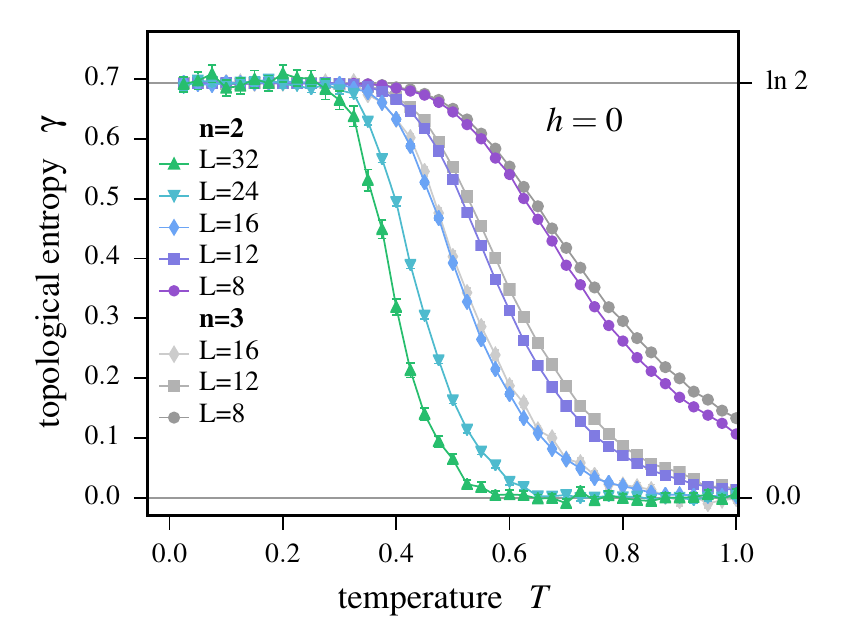}
\includegraphics[width=\linewidth]{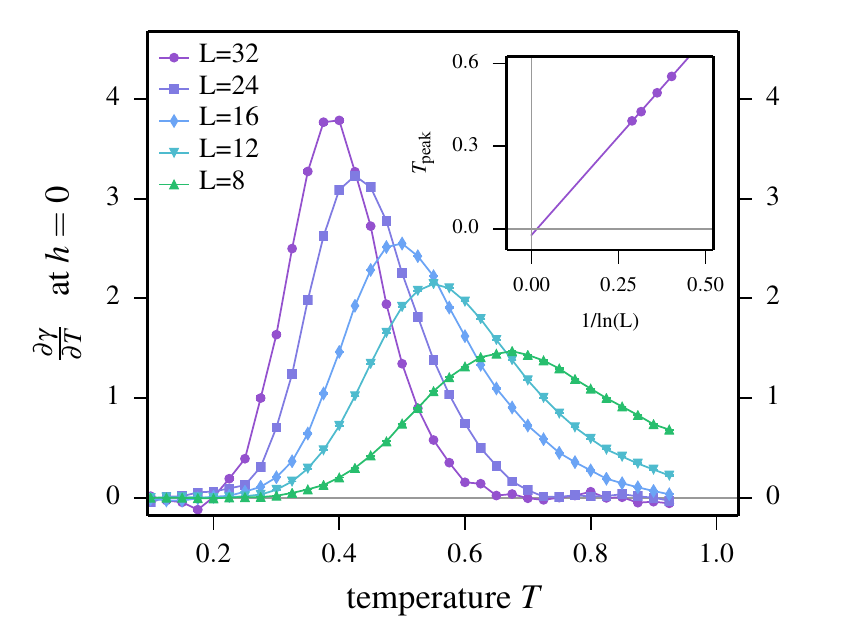}
\caption{(color online) 
{\it Top panel:} Finite-temperature behavior of topological entropy for the classical toric code model indicating the thermal transition
		       from the low-temperature topologically ordered phase to the high-temperature paramagnet for various system sizes.
{\it Bottom panel:} Identification of the transition temperature into the topologically trivial phase by determining the peak position of
			the derivative $\partial \gamma / \partial T$ for various system sizes. 
			The scaling of the so-determined transition temperatures shows the expected logarithmic scaling with system size
			indicating that topological order is unstable at finite temperature in the thermodynamic limit \cite{castelnovo07_2}.}
\label{fig:finite_zero}
\end{figure}

We verify this argument by numerically calculating the vanishing of the topological signature in the \Renyi entropy via finite-temperature simulations, initially in the absence of a magnetic field, i.e. $h=0$. 
Our results are depicted in Fig.~\ref{fig:finite_zero} and clearly show the expected behavior, i.e. a vanishing of the topological entropy for sufficiently large temperatures. To determine the location of the actual crossover temperature we identify the inflection points of the topological entropy, see the lower panel of Fig.~\ref{fig:finite_zero}. We find that the so-determined crossover temperatures show a perfect $1/\ln L$ scaling.

At this point, one might ponder why no  intermediate behavior appears for this transition (similar to the ones discussed before) given that the replica technique again puts the system effectively at $\beta$ in part $B$ and at $2 \beta$ in part $A$. A related question is whether the crossover temperature indicates whether subsystem $A$ or $B$ looses its  topological order. We checked the latter  by calculating the next higher \Renyi entropy $S_3(A)$, which puts part $A$ at temperature $3 \beta$. As illustrated in Fig. \ref{fig:finite_zero} the crossover temperatures does not move at all, thus indicating that it occurs only when subsystem $B$ undergoes order to the disorder transition. 

This behavior can be qualitatively understood from considering, for instance, what happens to the boundary constraints for bipartition $A_4$ of the Levin-Wen summation scheme above the crossover temperature $T_\text{co}$ (where subsystem $A$ is still topologically ordered while subsystem $B$ is not). Open loops can now traverse subsystem $A$ such that the original closed-loop constraints for the two boundaries melt into a single constraint for the total boundary of part $A$. As a consequence, not only the topological $\ln 2$-contribution of bipartition $A_4$ vanishes (cf. the table in Fig.~\ref{fig:LevinWen}) but also all connectivity contributions (which arise from a similar boundary counting argument). 

\subsubsection*{Finite magnetic field}

\begin{figure}
\includegraphics[width=\linewidth]{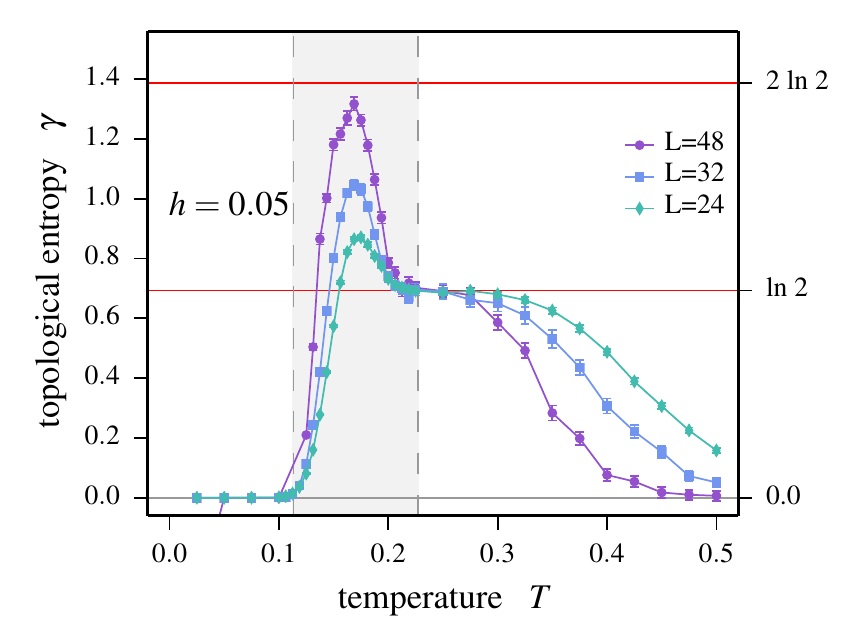}
\caption{(color online) 
		Finite-temperature behavior of $O(1)$ contribution to the \Renyi entropy $S^{(2)}$ of the classical toric code model 
		in the presence of a finite magnetic field/loop tension $h=0.05$. The $O(1)$ contribution exhibits the expected universal
		topological contribution $\gamma = \ln(2)$, which contributes for sufficiently large temperatures and vanishes at high
		temperatures.
        An additional connectivity contribution of $\gamma = \ln(2)$ is observed in an intermediate regime. It can be understood from noting that the \textit{effective} field is $h_\text{eff}=\beta h$ as we introduced $\beta$ in the Hamiltonian. Hence this field increases for $T \rightarrow 0$ and we observe the same behavior as in Fig.~\ref{fig:gamma_total} -- read from right to left. 
        The grey shaded area indicates the regime of $h_c/2 < h_\text{eff} < h_c$.
        At very low temperature even a small field $h=0.05$ leads to the frozen trivial phase so that no topological order is present and thus $\gamma=0$.
		}
\label{fig:finite_peak}
\end{figure}

Finally, we turn to the combined effect of finite-temperature defects and loop tension in the toric code.
For such a model the statistical weights of a loop configuration $\sigma$ becomes
\begin{equation}
  w(\sigma)=\exp\left(-\beta J_v D_\sigma + \beta h m_\sigma\right) \,,
    \label{eq:finfin}
\end{equation}
where $D_\sigma$ denotes the number of loop defects and $m_\sigma$ is the total magnetization.
Note that in contrast to the zero-temperature study above, we have now introduced an explicit temperature also in the weight stemming from a finite loop tension. 
In particular, this implies that if the temperature is too low compared to the loop tension $h$, there is no topological order since all fluctuations of the loop gas are frozen out  from the loop tension and the configuration is a single trivial state, see Fig. \ref{fig:finite_peak}. Only if $h \ll \beta^{-1} \ll J_v$, the system samples all loop configurations with approximately equal weight and topological order is observed. To this extent, $h_\text{eff} = \beta h$ takes the role of an effective loop tension. Therefore, the left half of Fig.~\ref{fig:finite_peak}  shows precisely the same physics as Fig. \ref{fig:gamma_total} seen from right to left -- including the occurrence of a connectivity contribution for an intermediate regime $h_c/2 < h_\text{eff} < h_c$ of the effective loop tension. On the other hand, for sufficiently high temperatures we again start to thermally activate defects resulting in a thermal crossover where the topological entropy drops to zero (see the right half of Fig. \ref{fig:finite_peak}). Again we see that this crossover temperature vanishes like $1/\ln L$ with increasing system size -- precisely mimicking the behavior seen for the system without magnetic field in Fig.~\ref{fig:finite_zero}.

%
%

\section{Discussion}
\label{sec:outlook}

To summarize, we have studied the breakdown of topological order in classical many-body systems from a \Renyi entropy perspective. One main result is the occurrence of an intermediate coupling regime, in which partial subsystem ordering leads to additional $O(1)$ contributions to the \Renyi entropy -- a scenario, which we extensively discussed in the context of the classical toric code model.

While we have concentrated our discussion on the Levin-Wen summation scheme \cite{LevinWen} to extract these $O(1)$ contributions it is probably worthwhile to point out that all observations reported here analogously apply to the alternative Kitaev-Preskill summation scheme \cite{KitaevPreskill} as well. We note in passing that in applying the  Kitaev-Preskill summation scheme to detect topological order in classical systems, one needs to (i) adapt a topologically non-trivial geometry for the bipartions (e.g. a donut geometry) in contrast to the application for quantum systems and (ii) be aware of the fact that not all corner contributions are fully eliminated in this scheme \cite{MariaPrivate}.

A natural question is to ask whether such effects may also occur in quantum systems. 
To this end, it is worth pointing out that the replica scheme to calculate the quantum \Renyi entropies is defined in the world-line representation of the partition function, i.e. replicas are stitched together in imaginary time or inverse temperature. Thus, the reported artifacts would only occur when considering {\em thermal} phase transitions of quantum systems and would not be expected to occur for zero-temperature quantum phase transitions  driven by some coupling parameter (which is not explicitly used in the replica representation).

On the other hand, we note that the classical \Renyi entropies we discussed here are directly related \cite{FootnoteShannonEntanglement} to the basis-dependent Shannon entanglement entropy introduced and studied in Refs.~\onlinecite{Stephan,Alcaraz13,SMI,Alcaraz14}. These basis-dependent Shannon and \Renyi entropies as well as related quantities have been shown to exhibit unexpected additional signatures when driving a system through a phase transition, reflecting the same physics discussed here. 

Taking an even wider perspective, we might think of our results as indications of the limitations of the \Renyi entropies with regard to their $n=1$ counterparts, i.e. the von-Neumann entanglement entropy and the classical Shannon entropy. Other limitations of the \Renyi entropies have been discussed earlier, e.g. in the context of the $O(N)$ model \cite{Metlitski} as well as Rokhsar-Kivelson type wave functions \cite{Stephan,SJFM,Chandran}, which both show additional (boundary) phase transitions for \Renyi indices $n\geq 2$.
In the latter case the quantum entropy may be mapped onto a classical problem with different replicas stitched together, but all at the same temperature. Such a gluing can be sufficient to trigger an ordering, not in the bulk but at the boundary. In both cases a precise renormalization group analysis is needed to address the possibility of an intermediate boundary phase transition. 
For generic, more complicated states such types of ordering may or may not occur -- the precise behavior as a function of  \Renyi index $n$ would (likely) need to be worked out case by case depending on the model and on spatial dimensionality. So far all known examples refer to critical states without topological order in spatial dimension $d>1$, which are driven away from criticality due to the gluing of several replicas. However, it appears unlikely that such a phenomenon could occur in gapped topologically ordered systems at zero temperature. Indeed in that case no connectivity contribution has been observed in numerical simulations so far \cite{IsakovHastingsMelko}, even though these are typically limited to relatively small system sizes.


\section*{Acknowledgments}

We thank P. Calabrese, M. Hermanns, and A. Rosch for insightful discussions. 
S.T. thanks R. G. Melko for an inspiring exchange that has led to the idea for the calculations in the manuscript at hand. 
J.H. acknowledges support from the Bonn-Cologne Graduate School of Physics and Astronomy.
Our numerical simulations were performed on the CHEOPS cluster at RRZK Cologne.

%
%

\appendix

\section{Replica loopgas sampling}
\label{app:sampling}

In this appendix we provide a detailed account of the adaptation of non-local loop updates to the replica setup. 

\begin{figure}[b]
  \centering
  \includegraphics[width=\linewidth]{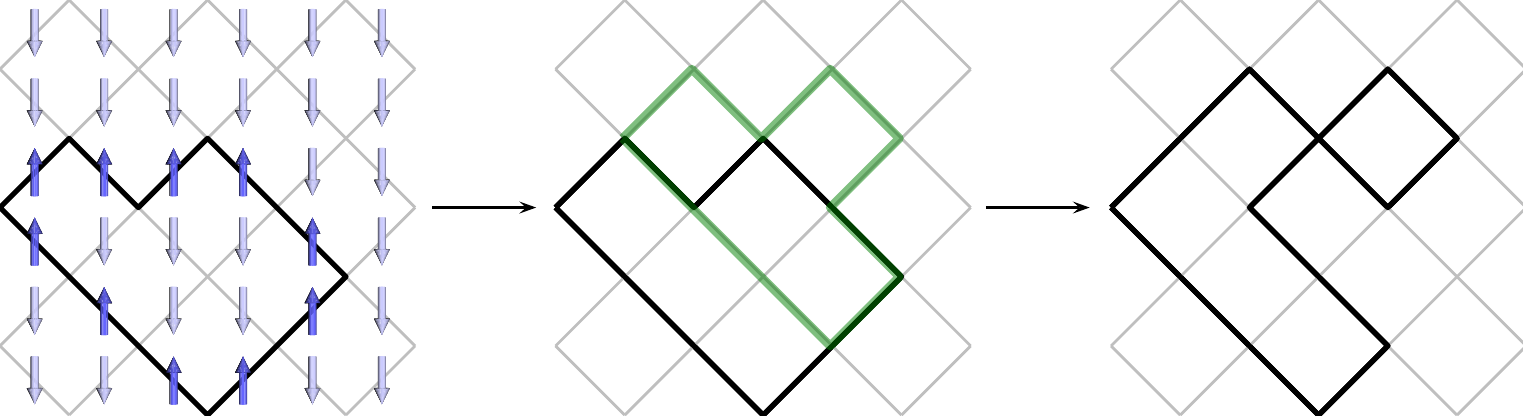}
  \caption{(color online) A loop update is performed by creating a template loop (green) in the lattice and applying it to the spin configuration with its spin loops (black) if the resulting energy change is accepted.}
  \label{fig:loopupdate}
\end{figure}

The core of a Monte Carlo simulation lies in the implementation of the Markov chain of configurations, i.e. the algorithm for selecting and accepting new configurations. This is sketched in the following. The major challenge when computing \Renyi entropies is that any configuration within the replica system has to satisfy the loopgas constraint in both replicas. Starting from a valid configuration, which could e.g. be the state with all spins pointing down, we will apply a modification to this state by creating or deleting loops. To do so we have the algorithm sample a \textit{template loop} -- a loop in the lattice which does not take care of the orientation of the spins. If accepted according to the detailed balance condition this template loop will be applied to the spins along this loop by flipping them. This procedure is shown in Fig. \ref{fig:loopupdate}. Sampling a template loop is particularly challenging if this loop traverses part $A$ since its degrees of freedom are identified with those in the other replica. Thus, two additional open loop ends are generated in the other replica at the boundary between $A$ and $B$. These two ends need to be linked in part $B$ of this other replica as well. Trying this, it is possible to traverse part $A$ again. Obviously this procedure can easily result in a seesaw of linking two open loop ends in part $B$ of either replica.

In order to generate the template loop a first approach is to perform a random walk through the lattice. Once a valid loop is found, we apply a Metropolis probability between the current and the potential next state to decide on its acceptance. This is discussed in \ref{sec:randomwalk}. We will then present in \ref{sec:directedwalk} a more advanced algorithm which unifies the selection and acceptance step in a so-called directed walk. Finally, the extension to finite temperatures is explained in \ref{sec:finitetemp}.

By a walk we mean a procedure that creates a series of adjacent vertices whose links carry a spin. During the algorithmic creation of the series the walk has a {\it head} which is the the last appended vertex. This head ``decides'' which vertex (and thereby which spin) to append next.

\subsection{Random walk}\label{sec:randomwalk}

\begin{figure}[t]
  \centering
  \includegraphics[width=\linewidth]{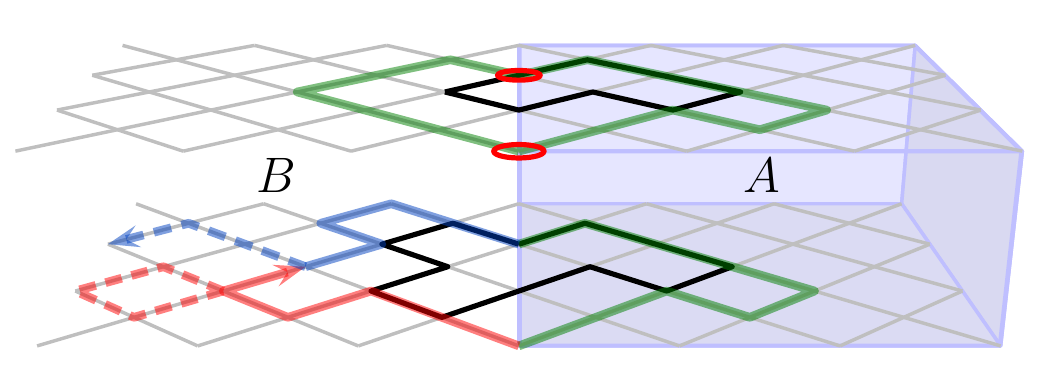}
  \caption{(color online) Visualization of the healing process. On top, a template loop (green) is sampled within the first replica. It leaves behind two open loop ends (red circles) at the boundary between $A$ and $B$ in the other replica. The right sketch depicts the healing processes: Two random walks (blue and red) simultaneously try to meet each other. The dashed segments indicate parts of the template loops which are discarded. Internal loops (dashed red) would be allowed but are contracted in order to shorten the total template loop. }
  \label{fig:discardhead}
\end{figure}

The random walk is started at an arbitrary vertex of the toric code lattice and proceeds with equal probability of $\frac14$ to one of the four neighboring vertices  \cite{FootnoteBouncing}.
Once a vertex is visited a second time, we can stop the walk and discard the first segment of the loop up to the first visit of this vertex. Possibly, the random walk has entered one or more times the connected subregion $A$ so that its segments in $A$ in the other replica also have to be included in the template loop. Thus, an even number of open template loop ends is created in the other replica at the boundary to part $B$. To match these ends, it is efficient to start individual random walks at every open end simultaneously. If the head of one of the walks hits a vertex already visited by a different walk, then we have created a loop that connects the two open ends, so that these two are {\it healed}. Typically, the two loops do not meet precisely at their heads so that there is a superfluous part between the meeting point and one of the heads. This part of the template loop is discarded, cf. Fig. \ref{fig:discardhead}. It may also happen, that a specific random walk head hits a vertex twice. In this case, we can discard the resulting {\it internal} loop of this walk since it does not help in linking open ends. Moreover, the general strategy is to keep the total length of the template loop as short as possible.

In Monte Carlo parlance, the procedure described above ensures ergodicity. What remains is to fulfill detailed balance. Since we have only generated a template loop so far, we can compare the total weight of the configuration prior to the application of this template loop and after it. In our special case we need to determine the magnetizations $m_\text{before}$ and $m_\text{after}$. The acceptance is then decided on by the Metropolis probability 
\begin{equation}
  p(\sigma \rightarrow \sigma~ \circ~ \text{template loop}) = \min (1, e^{h (m_\text{after}-m_\text{before})} ),
  \label{eq:detailedbalance}
\end{equation}
where the function composition symbol $\circ$ is used to denote the application of the template loop on the current configuration.

Since this implementation of the Markov chain separates the selection and acceptance of new configurations we can almost freely design the random walk procedure as long as we guarantee ergodicity. A shortcoming of it appears at higher loop tension $h$ where the acceptance probability of an energetically less favourable configuration is low. In addition, for relatively large sizes of the subsystem $A$, the healing of all open loop ends can entail many random walks back and forth between the replicas. This makes the algorithm less efficient, especially for the parameter settings ($h \approx 0.44$, $L$ large) we want to investigate to track the phase transitions.


\subsection{Directed walk}\label{sec:directedwalk}
A more efficient approach to the Monte Carlo update scheme is to unify selection and acceptance. The idea is to ensure detailed balance on-the-fly while generating the template loop. Strictly speaking it is no longer a {\it template} loop since all spins along this loop are flipped with probability 1, i.e. they can (and have to) be flipped directly. The walk that the head of the loop performs is not at random but obeys probability rules depending on the value of the potential next spins to be visited. In general, the rule is to select the next spin (which unambiguously selects the next vertex) with so-called heat bath probability. This means, the microscopic weights $w_i$ of all candidates for the next head of the loop are added up to a normalization constant $n$. In the toric code model the head has four possibilities to choose the next spin because bouncing must now be included. The only possible values for the weights are $w_i \in \{\exp(h), \exp(-h), \exp(2h), \exp(-2h)\}$, where the factor of two in the exponent applies in the connected part $A$ which is simulated at an effective loop tension of $2h$ as discussed above. With the constant $n= \sum_{i=1}^4 w_i$ the heat bath probabilities for the four directions are defined as
\begin{equation}
  \frac{w_1}{n}, \frac{w_2}{n}, \frac{w_3}{n}, \frac{w_4}{n}.
  \label{eq:heatbath}
\end{equation}

\begin{figure}[b]
  \centering
  \includegraphics[width=\linewidth]{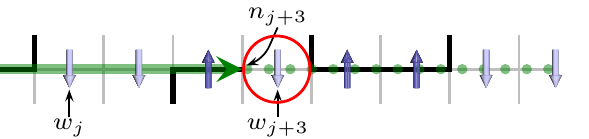}

  \vspace{10pt}
  \includegraphics[width=\linewidth]{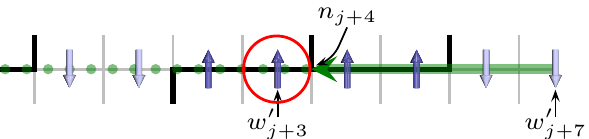}
  \caption{Formation of the probabilities of two reversing update loops at a specific intermediate position $j+3$ of the walk using heat bath probabilities. We compare the probabilities of selecting the encircled spin in the right moving loop and in its reversal left moving counterpart. The probability in the right moving case is $w_{j+3} / n_{j+3} = \exp(h) / (\exp(-h) + 3\exp(h))$. Note that spins at $j \dots j+2$ are already flipped when the loop does this selection at $n_{j+3}$ since we perform the update in-place. For the reverse move the selection probability is $w'_{j+3} / n_{j+4} = \exp(-h) / (3\exp(-h) + \exp(h))$. In particular, the ratio of these to probabilities is {\it not} $\exp(2h)$ as it should be if we wanted a microscopic detailed balance for the spin flip at $j+3$. }
  \label{fig:individualbalance}
\end{figure}

It is not obvious to see whether this rule generates an update which obeys the detailed balance condition. To check it we need to identify a reverse update to any (completed) update and compare the probabilities of their occurence. If we now try to fractionize this update into its individual moves according to the heat bath rule and compare it to its reverse move, cf. Fig. \ref{fig:individualbalance}, we see that the normalization constants thwart the analysis: A specific spin at position $j$ in the loop was chosen to be visited (and flipped) with probability $w_j/n_j$ in the original update. Doing its reverse, we would have the probability $w_j^{-1}/n'_{j+1}$ to reflip it. The ratio of the two probabilities is in general not $w_j / w_j^{-1}$ as they should be from the ratio of the weights of the configurations that differ by flipping spin $j$. Two facts about the involved normalization constants need to be dealt with: (i) they originate from different configurations and (ii) they are shifted by a lattice position. 

\begin{figure}[h!]
  \centering
  \includegraphics{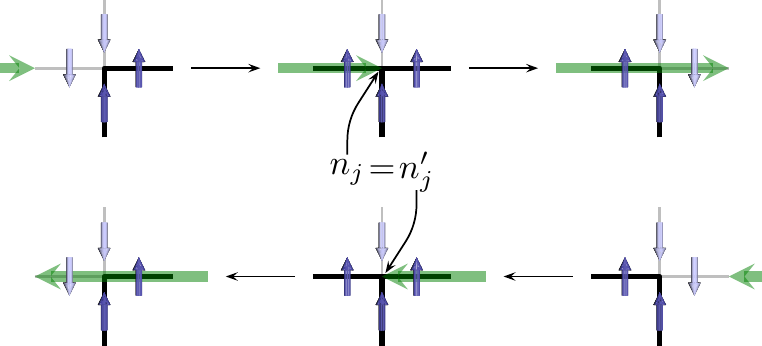}
  \caption{(color online) Passing of the directed walk (green) at vertex $j$. The normalization constants for inverse loop update walks agree. In this example they both have the value of $n_j = 3 \exp (-h) + \exp(h)$.}
  \label{fig:normalizconst}
\end{figure}

Issue (i) is not a problem since the value of the normalization constant at a specific vertex at the moment when it is the head is the same for both the original and its reverse update. This can be understood from Fig. \ref{fig:normalizconst} where we see that the coming-from and the going-to spin switch roles (and weights) in both situations.

The other issue (ii) can be resolved by leaving the local perspective from the loop head and regarding the entire loop update. If we consider a chain of moves and its reversion, we have to multiply all probabilities to obtain the total probability for the walk. In the ratio of the total probabilities for the walk and its reverse, we realize that all normalization constants vanish. Since their respective numerator-denominator pairing is shifted by one in the product, we dub this feature of the microscopic rule {\it staggered detailed balance}. The staggering can be seen in Fig. \ref{fig:individualbalance} and the ratio between a loop update and its reverse is given later in Eq. (\ref{eq:simpledb}).

We have seen that detailed balance can be satisfied in principle by the design of the walk but crucial aspects have not been discussed yet: How is a loop initiated and finished and what are the decision rules for choosing the next spin at the boundary between $A$ and $B$? It is clear that a loop must bifurcate when leaving part $A$ but how is a reunification at another site at the boundary accepted? The latter may fail such that the update is discarded for technical reasons (as opposed to probabilistic reasons).

For the start of a loop update, a random spin in a random replica is chosen and one of its adjacent vertices is selected with probability of $\frac{1}{2}$. This spin is immediately flipped and the walk continues at the selected vertices following the heat bath rule. It is instructive to note at this point that we have not ``paid'' the flipping of this first spin in terms of acceptance probabilities. This will be caught up at the decision to end the loop. The probability to start the loop at a particular pair of a spin and a vertex is thus simply
\begin{equation}
  p_\text{init} = \frac{1}{4N},
  \label{eq:pinit}
\end{equation}
where $N$ is the total number of spins in the lattice.

In order to finish the walk it is first of all necessary that the loop be closed, i.e. that the head reaches precisely the other vertex at the initial spin --- the vertex that was {\it not} chosen to start with. This is not sufficient, since just like anywhere the head is free at this vertex to chose any of the adjacent spins --- among them the initial spin. Before we determine the heat bath probabilities and normalization constant we have to flip the initial spin once again. The loop head thus ``sees'' the original orientation of this spin before the whole loop update was started. Afterwards the heat bath selection is performed. In the case, the inital spin is chosen, it is flipped a third time and the loop update is successfully finished. Only now, the flipping of the initial spin is ``paid'' by a probabilistic selection according to its weight. If another but the initial is chosen, the loop continues and we must not forget to flip the initial spin again. 

\begin{figure}[h!]
  \centering
  \includegraphics[width=\linewidth]{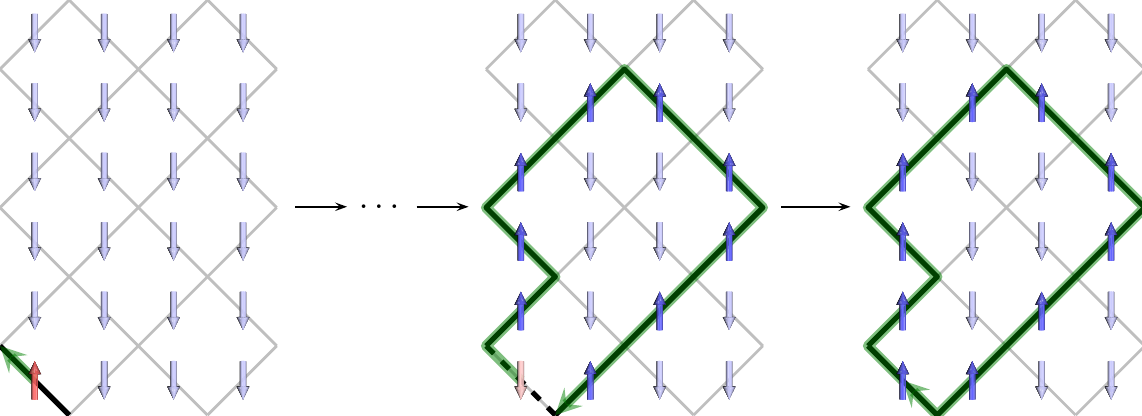}
  \caption{(color online) Completing a loop by catching up the probabilistic choice of the initial spin as part of the loop. The initial spin is drawn red in order to emphasize that it was not {\it paid} in terms of probabilistic weight selection until the last step. In the middle sketch it is temporarily flipped back since the loop head must choose it as if it was never flipped. }
  \label{fig:completeloop}
\end{figure}

We have now explained the algorithm for non-boundary-crossing loops. Fig. \ref{fig:completeloop} and the following equation prove that the detailed balance condition is fulfilled.
\begin{align}
     & \frac{p(\sigma \rightarrow \sigma')}{p(\sigma' \rightarrow \sigma)} \nonumber \\ 
   =~~ & \frac{%
    p_{\text{init},s_0} \cdot ~\frac{w_{s_1}}{n_1}~ \frac{w_{s_2}}{n_2}~ \dots ~\frac{w_{s_{l-1}}}{n_{l-1}}~ \frac{w_{s_0}}{n_l} \qquad\qquad%
    } {%
    \qquad\quad  \frac{w^{-1}_{s_0}}{n_1} \frac{w^{-1}_{s_{1}}}{n_{2}}~ \dots~ \frac{w^{-1}_{s_{l-2}}}{n_{l-1}} \frac{w^{-1}_{s_{l-1}}}{n_l} \cdot p_{\text{init},s_0} %
    } \nonumber\\ 
  =~~ & \frac{w(\sigma')}{w(\sigma)}.
  \label{eq:simpledb}
\end{align}
The relative shift of the numerators 
again demonstrates what we mean by staggered detailed balance for the loop steps. The chronological order of the appearance of the probabilities is from left to right in the main numerator but from right to left in the main denominator. We chose to write the equation in this way to emphasize the reversing effect of factors that are written one below the other.

It remains to define rules to deal with boundary crossings and the bifurcations of the loop that follow. In order to justify the probabilities for the direction of the walk we need to ensure reversibility of any moves we allow. This is the reason why we only allow a single bifurcation of the loop and hope for its reunification. In case the reunification fails, the update is aborted and the original configuration is restored. We will discuss this point further at the end of this section. The vertices of interest for the treatment of boundary crossings are those which have adjacent spins in either of the bipartitions. In the following we call them {\it boundary vertices}. In general we will never set a heat bath normalization constant at a vertex using weights from both parts, i.e. both $\exp(\pm h)$ and $\exp(\pm 2h)$. In other words, a drift to or away from the connected part $A$ is avoided.

\begin{figure}[h!]
  \centering
  \includegraphics[width=\linewidth]{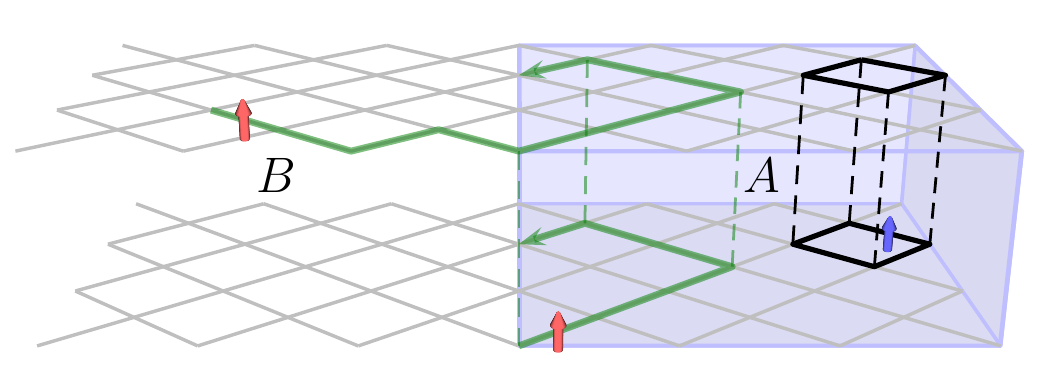}
  \caption{(color online) In the presence of a connected subsystem $A$ loops need to bifurcate. After entering part $A$, all weights have to be taken using the effective loop tension of $2h$ and flipped in both replicas. The first spin in part $A$ in the other at the entrance is unpaid (drawn red), just like the initial spin in part $B$. The bifurcation happens when the loop leaves part $A$ again and must be finished individually in both replicas by joining the red spins.}
  \label{fig:loopbifurcation}
\end{figure}

We first describe the update for the case of an initial spin somewhere in part $B$. If a boundary vertex is visited coming from the disconnected part $B$ we treat all four spins as if they were in part $B$ (although at least one is in part $A$) for the calculation of the heat bath weights. In other words, the head does not {\it see} the boundary. If the head selects a spin in part $A$, also its counterpart in the other replica is flipped. However, the flipping of this counterpart is at that state not paid in terms of acceptance probability just as we used this wording above. The loop continues in part $A$ where it uses the weights $\exp(\pm 2h)$ until it reaches again a boundary vertex. At such an occurrence we have to perform the mentioned bifurcation: First, a loop is continued in the initially chosen replica only. This loop should now rejoin the initial spin in the same way we described above for the simpler case devoid of a connected part $A$. Trying this, the loop must not enter part $A$ again. If it does, we have to abort the entire update. Second, another loop is continued in the respective other replica. In the same manner, this loop is expected to rejoin the by then unpaid spin in part $A$ at the entering boundary spin. To successfully finish the total loop update, this unification must happen without a prior visit of part $A$ again. Fig. \ref{fig:loopbifurcation} illustrates the described procedure.

A slight difference needs to be made if the initial spin of the update is in part $A$. In this case, the first visit of a boundary vertex leads to bifurcation and the bifurcated loop in replica $1$ of part $B$ is free to reenter part $A$ at any boundary vertex, creating there an unpaid boundary spin in replica $2$. Next, the other bifurcated loop in replica $2$ of part $B$ must walk until it reaches be performed and likewise go to this unpaid boundary spin that its counterpart has selected for reentering part $A$. Only if this succeeds, is the subsequent loop in part $A$ continued and then the update can be completed by going back to the initial spin.

Our implementation of the Monte Carlo algorithm does not allow multiple pairs of temporary open loops in part $B$ in one of the replicas. In principle one could allow more than one pair of open loops and perform the loop update in the initial replica until it rejoins the initial spin. Subsequently one could start to connect all $2n_o$ open ends in part $B$ of the other replica. Doing so it would be important to start the connecting walks only at those $n_o$ ends that arose from {\it leaving} part $A$ in the initial replica. Special care must be taken in this approach to not mix the order of the start spins of the healing walks in part $B$ of the other replica. They must be started in the same order the arose from the initial loop. This is due to the ambiguity of possible paths that lead to the same loop update. Mixing the order of starting spins would violate the detailed balance condition because the reversal loop of a loop could then be achieved through more than one move. 

We decided to allow only a single pair of bifurcations of the loop and avoid a long seesaw of healing processes for a single update step.


\subsection{Finite-temperature simulations}\label{sec:finitetemp}
So far, we have considered the classical toric code at zero temperature, i.e. the system subjected to the loopgas constraint. By releasing this constraint and activating the vertex terms in the Hamilton function (\ref{eq:hamiltonian}), we can investigate the toric code at finite temperature. In this situation pairs of open loop ends are permitted at the energetical cost of two times the vertex coefficient $J_v$. In practice we set $J_v=1$ since only the ratio between $J_v$ and $h$ is important. A loop update in our Monte Carlo simulation must therefore be able to introduce open loops. Once again we propose two strategies.

Our first approach consists in separating the operations of the loop update and the introduction of loop excitations. We pick two vertices of the total system at random and have the algorithm perform a random walk between them. At $h=0$ we only need to determine an acceptance probability for the open ends, which we set to Boltzmann weights $\exp(\pm J_v)$ or $\exp(\pm 2 J_v)$ depending on the subsystem where the open end is introduced. In case one of the vertices is in part $A$ and the other in part $B$ we need a third vertex in part $B$ of the respective other replica and thus create a ``double-tongued'' open loop. Also at finite loop tension we can apply Metropolis probabilities to the acceptance of a sampled open loop by balancing both the change in energy and in magnetization between the current and the proposed configuration.

\begin{figure}[t]
  \centering
  \includegraphics[width=\linewidth]{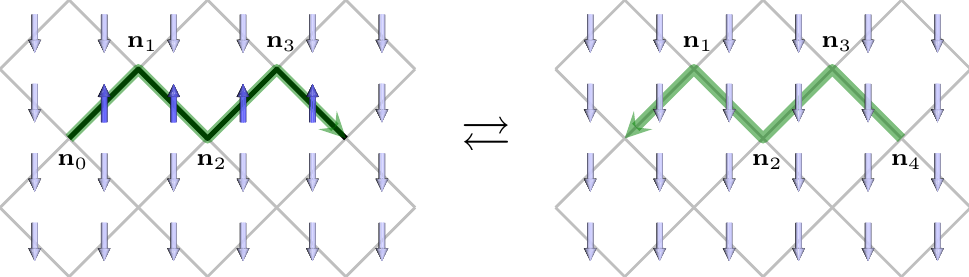}
  \caption{(color online) Two mutually reversing open loop updates including the normalization constants used for the respective choice of the next direction of the walk. The last normalization $n_4$ (resp. $n_0$) is missing and therefore will be included artificially in the decision to stop the loop.}
  \label{fig:edgeconstants}
\end{figure}

Going beyond the simple sampling technique we also present an algorithm that again includes the acceptance probability in the selection of the subsequent configuration. We therefore extend our zero energy algorithm based on heat bath selections for the random walk at every vertex in the lattice. The head of the loop may take four different directions at a specific vertex. We add a fifth event to these four, namely stopping the loop and thereby creating (or annihilating) an open loop end in the spin configuration. However, we will not include the decision of stopping the loop in the heat bath sum but perform it separately using the Boltzmann weight of an open loop end. This decision is made prior to the selection of the next bond to walk on. There is one issue concerning the detailed balance between an open loop update and its reversal update: Due to the staggered detailed balance philosophy the last heat bath normalization constant of the walk does not enter any of the probabilities for the loop continuation but the first one does, see Fig. \ref{fig:edgeconstants}. Roles are switched in the reversal update, which implies that the two normalization constants at the edges do not vanish in the ratio between the opposite loop updates.

Our workaround is to artificially include every heat bath normalization constant (and thus also the last one) into the decision of stopping the loop. The probability of stopping at the $j$ths vertex of the loop is thus $\frac{\exp(\pm J_v)}{n_j}$. Since in our model $n_j>1$ in any case, this alteration of the pure Boltzmann constant never leads to trivial probabilities ($>1$) for stopping the loop. In formulas the ratio between two opposing loop updates is given by

\begin{widetext}
\begin{equation}
      \frac{p(\sigma \rightarrow \sigma')}{p(\sigma' \rightarrow \sigma)}%
   =  \frac{%
       p_{\text{init},v_0} \cdot w_{v_0} ~\frac{w_{s_0}}{n_0}~(1-w_{v_1}) \frac{w_{s_1}}{n_1} (1-w_{v_2})~ \dots ~(1-w_{v_{l-1}})\frac{w_{s_{l-1}}}{n_{l-1}}~ \frac{w_{v_l}}{\text{\boldmath $n_l$}} \qquad\qquad%
    } {%
        \qquad\quad  \frac{w^{-1}_{v_0}}{\text{\boldmath $n_0$}}\frac{w^{-1}_{s_0}}{n_1} (1-w_{v_1})\frac{w^{-1}_{s_1}}{n_2}(1-w_{v_2})~ \dots~ (1-w_{v_{l-1}})\frac{w^{-1}_{s_{l-1}}}{n_l} \cdot w^{-1}_{v_l} \cdot p_{\text{init},v_l} %
    } %
  =  \frac{w(\sigma')}{w(\sigma)}.
  \label{eq:finitetemperaturedb}
\end{equation}
\end{widetext}

Until now we have neglected again the bipartition and replication of the lattice. Surprisingly, the situation becomes easier if open loops are allowed. We can simply forbid boundary crossings of open loops, i.e. abort any loop update which tries it. The algorithm is still ergodic since an arbitrary loop update (including a boundary crossing) can by generated by two non-crossing open loops which share one of their ends at the boundary. Moreover, also closed loops can be created that way by creating an open loop whose starting and ending vertex coincide. This does not make the closed loop algorithm superfluous since it is indispensable at the hard loopgas constraint and more efficient for very low temperature in combination with the open loop algorithm.

%
%


\begin{thebibliography}{99}
\bibitem{FootnoteAmbiguities}
While the topological entanglement entropy is a powerful indicator of topological order, it does not allow to fully identify the topological quantum field theory (TQFT) underlying this order. An example of such an ambiguity are the TQFTs for the toric code and the double semion theory, which both lead to a topological entropy of $\gamma = 2 \ln 2$.
\bibitem{FootnoteBoundaryConditions}
For completeness, it should be mentioned that this mapping is strictly defined for the even parity sector of the toric code model. Other parity sectors of the toric code would necessitate the definition of  antiperiodic boundary conditions in the Ising model for every odd winding number in the toric code model.
\bibitem{FootnoteShannonEntanglement}
This connection can be most easily seen for the mutual information of an infinite strip or cylinder, cut into two parts, and expressing the Shannon mutual information in the transfer matrix formalism (see e.g. Ref.~\onlinecite{GMI}), one is led to a particular basis that allows the definition of a Shannon entropy of the ground state. For general \Renyi index $n \geq 2$ such a relation can also be established.
\bibitem{FootnoteBouncing}
Except for the first step, the respective precursor in the walk can be excluded from the choice (avoid bouncing) so that the probability is $\frac13$ for the remaining directions.
\end{thebibliography}
\end{document}